\input harvmac
\let\includefigures=\iftrue
\let\useblackboard==\iftrue
\newfam\black

\includefigures
\message{If you do not have epsf.tex (to include figures),}
\message{change the option at the top of the tex file.}
\input epsf
\def\figin{\epsfcheck\figin}\def\figins{\epsfcheck\figins}
\def\epsfcheck{\ifx\epsfbox\UnDeFiNeD
\message{(NO epsf.tex, FIGURES WILL BE IGNORED)}
\gdef\figin##1{\vskip2in}\gdef\figins##1{\hskip.5in}
\else\message{(FIGURES WILL BE INCLUDED)}%
\gdef\figin##1{##1}\gdef\figins##1{##1}\fi}
\def\DefWarn#1{}
\def\figinsert{\goodbreak\midinsert}
\def\ifig#1#2#3{\DefWarn#1\xdef#1{fig.~\the\figno}
\writedef{#1\leftbracket fig.\noexpand~\the\figno}%
\figinsert\figin{\centerline{#3}}\medskip\centerline{\vbox{
\baselineskip12pt\advance\hsize by -1truein
\noindent\footnotefont{\bf Fig.~\the\figno:} #2}}
\endinsert\global\advance\figno by1}
\else
\def\ifig#1#2#3{\xdef#1{fig.~\the\figno}
\writedef{#1\leftbracket fig.\noexpand~\the\figno}%
\global\advance\figno by1} \fi

\def\id{{1 \kern-.28em {\rm l}}}

\def\Z{{\bf Z}}

\def\K3{{\bf K3}}
\def\journal#1&#2(#3){\unskip, \sl #1\ \bf #2 \rm(19#3) }
\def\andjournal#1&#2(#3){\sl #1~\bf #2 \rm (19#3) }

\def\bar{\overline}
\def\hat{\widehat}
\def\ie{{\it i.e.}}
\def\eg{{\it e.g.}}

\def\frac#1#2{{#1\over#2}}

\def\half{\frac12}

\def\inbar{\,\vrule height1.5ex width.4pt depth0pt}
\def\IC{\relax\hbox{$\inbar\kern-.3em{\rm C}$}}
\def\IR{\relax{\rm I\kern-.18em R}}
\def\IP{\relax{\rm I\kern-.18em P}}
\def\Z{{\bf Z}}

%
%

%
\catcode`\@=11
\def\slash#1{\mathord{\mathpalette\c@ncel{#1}}}
\overfullrule=0pt

\def\underrel#1\over#2{\mathrel{\mathop{\kern\z@#1}\limits_{#2}}}

\catcode`\@=12


%

\def\tr{{\rm tr}}

\def\exp{{\rm exp}}


\def\ie{{\it i.e.}}
\def\eg{{\it e.g.}}


\lref\Kutasov{
D.~Kutasov,
``Two-dimensional QCD coupled to adjoint matter and string theory,''
Nucl. Phys. B {\bf 414}, 33-52 (1994)
doi:10.1016/0550-3213(94)90420-0
[arXiv:hep-th/9306013 [hep-th]].
}

\lref\Hornbostell{
K.~Hornbostel,
``THE APPLICATION OF LIGHT CONE QUANTIZATION TO QUANTUM CHROMODYNAMICS IN (1+1)-DIMENSIONS,''
SLAC-0333.
}

\lref\Hamer{
C.~J.~Hamer,
``Lattice Model Calculations for SU(2) Yang-Mills Theory in (1+1)-Dimensions,''
Nucl. Phys. B {\bf 121}, 159 (1977)
doi:10.1016/0550-3213(77)90334-0
}

\lref\ZubovURA{
  R.~A.~Zubov, S.~A.~Paston and E.~V.~Prokhvatilov,
  ``Exact solution of the 't Hooft equation in the limit of heavy quarks with unequal masses,''
Theor.\ Math.\ Phys.\  {\bf 184}, no. 3, 1281 (2015), [Teor.\ Mat.\ Fiz.\  {\bf 184}, no. 3, 449 (2015)].
}

\lref\Ziyatdinov{
I.~Ziyatdinov,
``Asymptotic properties of mass spectrum in 't Hooft's model of mesons,''
Int.\ J.\ Mod.\ Phys.\ A\ {\bf 25}, 3899-3910 (2010)
doi:10.1142/S0217751X10050287
[arXiv:1003.4304 [hep-th]].
}

\lref\Fonseca{
P.~Fonseca and A.~Zamolodchikov,
``Ising field theory in a magnetic field: Analytic properties of the free energy,''
[arXiv:hep-th/0112167 [hep-th]].
}

\lref\GrossM{
D.~J.~Gross, I.~R.~Klebanov, A.~V.~Matytsin and A.~V.~Smilga,
``Screening versus confinement in (1+1)-dimensions,''
Nucl. Phys. B {\bf 461}, 109-130 (1996)
doi:10.1016/0550-3213(95)00655-9
[arXiv:hep-th/9511104 [hep-th]].
}

\lref\KutasovV{
D.~Kutasov and A.~Schwimmer,
``Universality in two-dimensional gauge theory,''
Nucl. Phys. B {\bf 442}, 447-460 (1995)
doi:10.1016/0550-3213(95)00106-3
[arXiv:hep-th/9501024 [hep-th]].
}

\lref\tHooft{
G.~'t Hooft,
``A Planar Diagram Theory for Strong Interactions,''
Nucl. Phys. B {\bf 72}, 461 (1974)
doi:10.1016/0550-3213(74)90154-0
}

\lref\tHooftt{
G.~'t Hooft,
``A Two-Dimensional Model for Mesons,''
Nucl. Phys. B {\bf 75}, 461-470 (1974)
doi:10.1016/0550-3213(74)90088-1
}

\lref\Dalley{
S.~Dalley and I.~R.~Klebanov,
``String spectrum of (1+1)-dimensional large N QCD with adjoint matter,''
Phys. Rev. D {\bf 47}, 2517-2527 (1993)
doi:10.1103/PhysRevD.47.2517
[arXiv:hep-th/9209049 [hep-th]].
}

\lref\Bhanot{
G.~Bhanot, K.~Demeterfi and I.~R.~Klebanov,
``(1+1)-dimensional large N QCD coupled to adjoint fermions,''
Phys. Rev. D {\bf 48}, 4980-4990 (1993)
doi:10.1103/PhysRevD.48.4980
[arXiv:hep-th/9307111 [hep-th]].
}

\lref\Dempsey{
R.~Dempsey, I.~R.~Klebanov, L.~L.~Lin and S.~S.~Pufu,
``Adjoint Majorana QCD$_2$ at Finite $N$,''
[arXiv:2210.10895 [hep-th]].
}

\lref\DempseyY{
R.~Dempsey, I.~R.~Klebanov and S.~S.~Pufu,
``Exact symmetries and threshold states in two-dimensional models for QCD,''
JHEP {\bf 10}, 096 (2021)
doi:10.1007/JHEP10(2021)096
[arXiv:2101.05432 [hep-th]].
}

\lref\Katz{
E.~Katz, G.~Marques Tavares and Y.~Xu,
``Solving 2D QCD with an adjoint fermion analytically,''
JHEP {\bf 05}, 143 (2014)
doi:10.1007/JHEP05(2014)143
[arXiv:1308.4980 [hep-th]].
}

\lref\Dubovsky{
S.~Dubovsky,
``A Simple Worldsheet Black Hole,''
JHEP {\bf 07}, 011 (2018)
doi:10.1007/JHEP07(2018)011
[arXiv:1803.00577 [hep-th]].
}

\lref\Donahue{
J.~C.~Donahue and S.~Dubovsky,
``Classical Integrability of the Zigzag Model,''
Phys. Rev. D {\bf 102}, no.2, 026005 (2020)
doi:10.1103/PhysRevD.102.026005
[arXiv:1912.08885 [hep-th]].
}

\lref\Donahuee{
J.~C.~Donahue and S.~Dubovsky,
``Confining Strings, Infinite Statistics and Integrability,''
Phys. Rev. D {\bf 101}, no.8, 081901 (2020)
doi:10.1103/PhysRevD.101.081901
[arXiv:1907.07799 [hep-th]].
}

\lref\Kogut{
J.~B.~Kogut and L.~Susskind,
``Hamiltonian Formulation of Wilson's Lattice Gauge Theories,''
Phys. Rev. D {\bf 11}, 395-408 (1975)
doi:10.1103/PhysRevD.11.395
}

\lref\Toda{
Toda, Morikazu,
``Vibration of a Chain with Nonlinear Interaction."
 Journal of the Physical Society of Japan {\bf 22} (1967): 431-436.
 }

\lref\Hornbostel{
K.~Hornbostel, S.~J.~Brodsky and H.~C.~Pauli,
``Light Cone Quantized QCD in (1+1)-Dimensions,''
Phys. Rev. D {\bf 41}, 3814 (1990)
doi:10.1103/PhysRevD.41.3814
}

\lref\Pauli{
H.~C.~Pauli and S.~J.~Brodsky,
``Solving Field Theory in One Space One Time Dimension,''
Phys. Rev. D {\bf 32}, 1993 (1985)
doi:10.1103/PhysRevD.32.1993
}

\lref\Fitzpatrick{
A.~L.~Fitzpatrick, J.~Kaplan, E.~Katz and L.~Randall,
``Decoupling of High Dimension Operators from the Low Energy Sector in Holographic Models,''
[arXiv:1304.3458 [hep-th]].
}

\lref\Andreev{
O.~Andreev,
``Some Aspects of Three-Quark Potentials,''
Phys. Rev. D {\bf 93}, no.10, 105014 (2016)
doi:10.1103/PhysRevD.93.105014
[arXiv:1511.03484 [hep-ph]].
}

\lref\Bali{
G.~S.~Bali,
``QCD forces and heavy quark bound states,''
Phys. Rept. {\bf 343}, 1-136 (2001)
doi:10.1016/S0370-1573(00)00079-X
[arXiv:hep-ph/0001312 [hep-ph]].
}

\lref\Cornwall{
J.~M.~Cornwall,
``What Is the Relativistic Generalization of a Linearly Rising Potential?,''
Nucl. Phys. B {\bf128}, 75-92 (1977)
doi:10.1016/0550-3213(77)90301-7
}

\lref\Hortacsu{
M.~Hortacsu,
``Heun Functions and Some of Their Applications in Physics,''
doi:10.1142/9789814417532\_0002
[arXiv:1101.0471 [math-ph]].
}

\lref\Dhar{
A.~Dhar, G.~Mandal and S.~R.~Wadia,
``String field theory of two-dimensional QCD: A Realization of W(infinity) algebra,''
Phys. Lett. B {\bf 329}, 15-26 (1994)
doi:10.1016/0370-2693(94)90511-8
[arXiv:hep-th/9403050 [hep-th]].
}

\lref\Eichten{
Eichten, E. and Gottfried, K. and Kinoshita, T. and Lane, K. D. and Yan, T. -M.,
``Charmonium: The model,"
Phys. Rev. D {\bf 17}, no.11, 3090--3117 (1978)
10.1103/PhysRevD.17.3090
}

\lref\Gottfried{
Eichten, E. and Gottfried, K. and Kinoshita, T. and Kogut, J. and Lane, K. D. and Yan, T. -M.,
``Spectrum of Charmed Quark-Antiquark Bound States,"
Phys. Rev. Lett. {\bf 34}, no.6, 369--372 (1975)
10.1103/PhysRevLett.34.369
}

\Title{} {\centerline{(1+1)D QCD with heavy adjoint quarks }}

\bigskip
\centerline{\it Meseret Asrat}
\smallskip
\centerline{International Center for Theoretical Sciences}
\centerline{Tata Institute of Fundamental Research
} \centerline{Bengaluru, KA 560089, India}

\smallskip

\vglue .3cm

\bigskip

\let\includefigures=\iftrue
\bigskip
\noindent

In this paper, we determine at weak coupling the non--relativistic $n$--body Schr\"{o}dinger equation that describes the low--lying color singlet bound states of two dimensional adjoint $QCD$ with heavy quarks. In the case of three adjoint quarks, we show that the three--body equation reduces equivalently to the Schr\"{o}dinger equation that describes a point electric dipole in an electric field in a plane angular sector. We conjecture that the three--body problem is solvable. We show that the eigenstates are given in terms of the triconfluent Heun functions. Our conjecture implies that a bound state of three adjoint quarks is described by a particle confined in a two dimensional Cornell potential. We expect the $n$--parton problem also to be solvable in a similar approach. 

\bigskip

\Date{11/22}

\newsec{Introduction}

Quantum chromodynamics ($QCD$) is the fundamental theory that describes quarks and gluons (in four spacetime dimensions). In particle (accelerators and) detectors, the quarks and gluons are always observed bound together into hadrons. Thus, at low energy, the theory is believed to exhibit confinement. The main goal in $QCD$ (and in general in Yang--Mills ($YM$) theory) has been to understand confinement (and/or the existence of a mass gap). However, a complete understanding of the phenomenon is still missing. In part this is because the phenomenon is non--perturbative and the theory is in general complex, for example, in terms of the number of dynamical degrees of freedom it contains and the phenomena it describes.

 In two spacetime dimensions, adjoint $QCD$ is a relatively simple and tractable theory that exhibits, among some other common properties, confinement (and at finite temperature deconfinement \Kutasov).\foot{Two dimensional fundamental $QCD$ does not exhibit a deconfinement transiton \Kutasov. } Therefore, it is useful to study this simple model to gain insights into confinement and other essential phenomena. A better understanding of the theory will be also useful in constructing a string worldsheet realization of $QCD$ strings. It is believed that at low energy the properties of $QCD$ might be reproduced by an effective theory of interacting long strings \refs{\Kogut,\ \Dubovsky}. In this paper, we consider this model with these perspectives in mind.

In two spacetime dimensions, a gluon has no physical propagating degrees of freedom since there are no transverse spatial dimensions. Therefore, it cannot form a color singlet bound state with a matter quanta. In adjoint $QCD$, thus, the quantum states are color singlet states of adjoint quarks bound together by non--dynamical, string--like, color flux tubes that confine the color gauge potential lines. The color singlet or gauge invariant bound states can contain two or more number of adjoint quarks. Thus, a color singlet bound state can be viewed as a chain of adjoint quarks on a closed string. However, depending on whether the number of the adjoint quarks is even or odd, the bound state is either a bosonic or fermionic state.

In two dimensional $QCD$ with fundamental quarks, all the meson states consist a quark and an anti--quark pair, and they are arranged in a single Regge trajectory \tHooftt. In adjoint $QCD$, on the other hand, it is expected that the states are grouped into separate multiple Regge trajectories \Dalley. See also \Kutasov. 

As suggested by `t Hooft \tHooft, considering the large $N$ limit (where $N$ is the rank of the gauge group) simplifies the theory. In this limit, there exists a systematic expansion in powers of $1/N$. This is easy to see since, in general, the theory can be obtained by dimensional reduction from higher dimensional gauge theories \refs{\Dalley,\ \Bhanot}. The theory, however,  despite being two dimensional and/or relatively simple, in the sense that the gluons are, for example, non--dynamical, has not been solved completely, even in the large $N$ limit. This is mainly because, in this limit, pair production and pair annihilation are not suppressed \refs{\Dalley,\ \Bhanot}. Therefore, the Hamiltonian relates states with different number of adjoint quarks or partons. This makes the computation of the exact spectrum analytically difficult. 

The adjoint spectrum has been computed, however, approximately in the large $N$ limit in \refs{\Dalley,\ \Bhanot} and recently for finite values of $N$ in \Dempsey. In these papers, the authors use discrete light--cone quantization and they numerically diagonalize the light--cone Hamiltonian. In this approach, the light--cone momentum and the momentum fraction carried by a quark are discretized. Thus, the (approximate) truncated space of states is finite dimensional and therefore diagonalizing the mass matrix is relatively tractable. More recently, the low energy approximate spectrum has been also computed by diagonalizing the Hamiltonian in a set of states created by operators with dimensions below a certain cut--off \Katz. The main point is that the two point functions of low dimension operators with a high dimension operator goes to zero exponentially fast. Therefore, the high dimension operators decouple from the low mass spectrum \Fitzpatrick. As a result, they can be ignored in the approximation with a small error. The error depends on the cut--off. Also more recently, a candidate relativistic Hamiltonian describing the high energy asymptotics of confining string has been obtained from effective long string worldseet theory \refs{\Donahue,\ \Donahuee,\ \Dubovsky}. The Hamiltonian equivalently describes a one dimensional chain of ordered massless particles with nearest neighbor interaction. The interaction potential is related to the potential in Toda lattice (in certain limit) \Toda, and the Hamiltonian has been shown to be super--integrable. 

The low--lying bound states of heavy quarks are believed to be described by a non--relativistic Schr\"{o}dinger equation. In this paper, we determine the non--relativistic Schr\"{o}dinger equation that describes the low--lying color singlet bound states of the two dimensional adjoint $QCD$ with heavy quarks. We work in the large $N$ or planar limit.  We keep the (`t Hooft) coupling parametrically small and fixed. We use the method employed in the papers \refs{\Hornbostell\Hamer\ZubovURA\Ziyatdinov -  \Fonseca}. In the paper \ZubovURA, the authors obtained at weak coupling the non--relativistic Schr\"{o}dinger equation that describes the `t Hooft model \tHooftt\ in the limit of heavy quarks and large number of colors. In this model the quarks are in the fundamental representation of the gauge group. They also computed (at weak coupling) exactly the eigenstates and the spectrum. Interestingly, the non--relativistic limit of `t Hooft model was actually discussed and same results were obtained earlier in \refs{\Hornbostell,\ \Hamer}.\foot{I thank Igor Klebanov for bringing to my attention these interesting earlier works.} 

In section two, we review in detail the method discussed in the papers \refs{\Hornbostell\Hamer\ZubovURA\Ziyatdinov -  \Fonseca}. We also discuss the results obtained in the papers \refs{\Hornbostell\Hamer - \ZubovURA} by applying the method to the two dimensional `t Hooft model \tHooftt. In section three, using the same method, we derive, at weak coupling, the equation that describes the low--lying bound states of the two dimensional adjoint $QCD$ with heavy quarks. We find that the equation equivalently describes a particle confined to the surface of an inverted $t$--gonal pyramid potential in $n$ dimensions. For a bound state with three constituent quarks we conjecture that the corresponding equation is exactly solvable.\foot{In the sense that one can write down a closed analytic expression.} We show that the eigenstates are given in terms of the triconfluent Heun functions. We discuss our approach and the spectrum of the bound states of two and three adjoint quarks in section four. On general grounds, we expect the $n$--body problem also to be solvable in a similar approach. 

We provided in appendix A representative plots of closed periodic orbits in the associated classical system of the three quarks system. We note that the classical dynamics is sensitive to initial conditions. On general grounds, we also expect sensitivity to initial conditions in the general case. Chaotic dynamical systems are in particular known to exhibit such behavior. However, in general, sensitivity to initial conditions alone does not necessarily imply chaos. Thus, the general $n$--body classical system might be of interest to gain insights into chaos theory.  In appendix B we collected some interesting intermediate results and useful equivalence relations.

\newsec{The large mass limit of the `t Hooft model}

In this section, we summarize the facts about the `t Hooft model of two dimensional $QCD$ \refs{\tHooft,\ \tHooftt} with gauge group $U(N)$ and fundamental fermions in the large constituent quark mass limit. In the next sections we will generalize this discussion to the case of two dimensional adjoint $QCD$. We will use the discussion presented in \ZubovURA, but we will take here the quark masses to be equal, $m_1=m_2=m$. See also \Hornbostell\ for a similar discussion.

The `t Hooft equation \refs{\tHooft,\ \tHooftt} involves the wavefunction of a meson (a bound state of quark and anti--quark pair), $\phi(\xi)$. Here $0\le \xi\le 1$ is the fraction of the light--cone momentum carried by one of the two quarks in the meson. Of course, the fraction carried by the other is $1-\xi$. The equation takes the form 
\eqn\aaaa{\mu^2\phi(\xi)=\alpha\left({1\over\xi}+{1\over 1-\xi}\right)\phi(\xi)-P\int_0^1d\xi'{\phi(\xi')\over(\xi'-\xi)^2},
}
where $\mu$ is a dimensionless\foot{The gauge coupling in two dimensions is dimensionful.} measure of the meson mass $M$,
\eqn\bbbb{M^2={g^2N\over\pi}\mu^2,
}
and 
\eqn\cccc{\alpha={\pi m^2\over g^2N}-1,
}
is a dimensionless measure of the size of the `t Hooft coupling, or equivalently the size of the coupling at the scale of the quark mass $m$.
Large $\alpha$ corresponds to weak coupling. $P$ in \aaaa\ stands for principal value (see \refs{\Hornbostell,\ \ZubovURA}). 

We are interested in studying this system in the limit $\alpha\gg 1$. Loosely speaking, the first term gives a large contribution, of order $\alpha$, to $\mu^2$, and the second term gives a small correction. Also, the first term can be thought of as the contribution of the masses of the quarks to the mass of the meson. For $g=0$, the second term, which is what gives confinement, is absent, and we get a continuum of values of $\mu^2$, starting from the minimal value obtained when $\xi=\half$, 
\eqn\ffff{\mu_0^2=4\alpha,
}
or using \bbbb, \cccc, $M^2=(2m)^2$. This is precisely what one would expect for a state of two quarks of mass $m$. As $\xi$ deviates from $\half$, the order $\alpha$ contribution to $\mu^2$ grows. Thus, if we want $\mu^2$ to be $4\alpha$ plus a small correction, we want the wavefunction $\phi(\xi)$ to be sharply peaked around $\xi=\half$ ( see also \Hornbostell).

Now, suppose we want to turn on the coupling $g$, while keeping the ratio $\alpha$ very large. In the notation of \ZubovURA, we take $a_1=a_2=1$, so $\alpha_1=\alpha_2=\alpha$, $k_1=k_2=\half$, and write 
\eqn\dddd{\xi=\half+\omega.
}
The `t Hooft equation \aaaa\ takes the form 
\eqn\eeee{\mu^2\phi(\omega)=\alpha\left({1\over\half+\omega}+{1\over \half-\omega}\right)\phi(\omega)-P\int_{-\half}^\half d\omega'{\phi(\omega')\over(\omega'-\omega)^2}.
}
As mentioned above, we are looking for states whose $\mu^2$ is of the form 
\eqn\gggg{\mu^2=\mu_0^2+\gamma,
}
where $\mu_0$ is given by \ffff, and $\gamma$ grows slower than $\alpha$ at large $\alpha$, \ie\ $\lim_{\alpha\to\infty}{\gamma\over\alpha}=0$. Substituting \gggg\ into \eeee, we get a `t Hooft type equation for $\gamma$. 
\eqn\hhhh{\gamma\phi(\omega)={4\alpha\omega^2\over {1\over4}-\omega^2}\phi(\omega)-P\int_{-\half}^\half d\omega'{\phi(\omega')\over(\omega'-\omega)^2}.
}
We are looking for solutions to this equation in which $\gamma\ll\alpha$. This means that the wavefunction $\phi(w)$ is sharply peaked around $w=0$. Thus, we can neglect the $\omega^2$ in the denominator on the r.h.s. of \hhhh, so it takes the form 
\eqn\iiii{\gamma\phi(\omega)=16\alpha\omega^2\phi(\omega)-P\int_{-\half}^\half d\omega'{\phi(\omega')\over(\omega'-\omega)^2}.
}
To formalize the requirement that for large $\alpha$, the wavefunction $\phi(\omega)$ is sharply peaked at $\omega=0$, we demand that if we rescale $\omega$ by a factor $t=t(\alpha)$, that we need to determine, \ie\ we write 
\eqn\dldl{\omega=st,
} 
then the wavefunction  
\eqn\jjjj{\phi(\omega)=\phi(st)=f(s),
}
where $f$ is a function that is not sensitive to $\alpha$. Plugging this ans\"{a}tz into \iiii\ and demanding that the two terms on the r.h.s. scale in the same way with $\alpha$ as $\alpha\to\infty$, we find that we must take 
\eqn\kkkk{t=\alpha^{-{1\over3}},
}
and, if we take this value for $t$, then $\gamma$ on the l.h.s. behaves like $\gamma\sim\alpha^{1\over3}$. Thus, it is convenient to define 
\eqn\llll{\gamma=\bar\gamma \alpha^{1\over3},
}
in terms of which the `t Hooft equation \iiii\ takes the form 
\eqn\mmmm{\bar\gamma f(s)=16s^2f(s)-P\int_{-\infty}^\infty ds'{f(s')\over(s'-s)^2}.
}
A number of things to note at this point:
\item{(1)} Since $\bar\gamma$ is obtained by solving a problem, \mmmm, which does not contain the expansion parameter $\alpha$, it does not depend on $\alpha$. Therefore, the solution for $\gamma$, \llll,  grows slower with $\alpha$ than the leading term in \gggg, in agreement with the assumptions that went into the analysis. 

\item{(2)} In going from \iiii\ to \mmmm\ we extended the range of integration. In fact, the correct range of integration in \mmmm\ should have been taken to be $-1/2t$ to $+1/2t$, with $t$ given by \kkkk.  In the limit $\alpha\to\infty$,  the boundaries of the integral go to infinity, so we expect the mistake in extending them to be small. How small depends on the behavior of the solution $f(s)$ for large values of the integrand. We comment on this later in the section.

The variable $s$ in \mmmm\ is a momentum type variable -- it is related via \dddd, \dldl, \kkkk, to the light--cone momentum fraction $\xi$ carried by a quark.\foot{One can think of $s$ as follows. In the c.o.m. frame, the two quarks have energy $E$ and momentum $\pm p$. $s$ is proportional to $p$, and the wavefunction $f(s)$ is the momentum space wavefunction of the bound state.} 
To solve \mmmm, it is useful to Fourier transform it to position space, as done in \ZubovURA:
we define 
\eqn\nnnn{\hat f(x)= {1\over2\pi}\int_{-\infty}^\infty ds f(s) e^{isx},
}
and write \mmmm\ as an equation for $\hat f(x)$,
\eqn\oooo{\bar\gamma\hat f(x)=-16{\hat f}\;''(x)+\pi|x|\hat f(x).
} 
The l.h.s. and the first term on the r.h.s. are obvious, and the second term on the r.h.s. relies on the definition of the principal value (see \eg\ equation (4) in \ZubovURA\ and equation (3.57) in \Hornbostell). 

Comments:

\item{(1)} Equation \oooo\ is interesting: it is the Schr\"{o}dinger equation for a particle in the potential $|x|$.\foot{This is also the equation that governs a point electric dipole on a line with electric field proportional to $x$. As we will see in section (4), viewing it in this picture is more useful.} This is basically the confining Coulomb potential in one spatial dimension. An interesting fact is that the treatment of the pole at zero momentum exchange in \mmmm\ (the $i\epsilon$ prescription associated with the principal value in that equation) is directly related to the fact that the potential rises both for positive and for negative $x$. 

\item{(2)} Of course, the momentum $s$ is light--like momentum, and the conjugate position variable $x$ is thus 
light--cone separation of the two quarks. Nevertheless, we get a compelling picture of the meson as a pair of quarks separated by the amount $x$ in a light--like direction, with the energy of the pair growing linearly with their separation. We will make use of this picture later, in the adjoint case. 

The solution of \oooo\ is an Airy function \ZubovURA. This is easy to see as follows. Consider first the region $x>0$. In this region, the Schr\"{o}dinger equation \oooo\ can be written as
\eqn\rrrr{\hat f(x)=g(y),
}
where $g(y)$ is a solution of the equation 
\eqn\pppp{g''(y)=yg(y),
}
and 
\eqn\qqqq{y= a(x - b), \quad a = \left({\pi\over 16}\right)^{1\over 3}, \quad b = {\bar\gamma\over \pi}.
}
This is in agreement with eq. (18), (19) in \ZubovURA\ and eq. (3.59), (3.60) in \Hornbostell. 

The solution of \pppp\ is $g(y)={\rm Ai}(y)$. The reason we need the ${\rm Ai}$ Airy function rather than the ${\rm Bi}$ is the usual: we need the solution to go to zero as $x,y\to\infty$, and the ${\rm Ai}$ function indeed goes to zero at infinity, while ${\rm Bi}$ blows up exponentially. 

Thus, for $x>0$ the solution to the Schr\"{o}dinger equation \oooo\ is $\hat f(x)={\rm Ai}(y)$. What about negative $x$? Since the problem \oooo\ is symmetric under $x\to -x$, there are two kinds of eigenstates, symmetric and antisymmetric under $x\to -x$. As usual, we will label the bound states by an integer $n$, with $n=0,2,4,\cdots$ corresponding to the symmetric solutions, and $n=1,3,5,\cdots$ corresponding to the antisymmetric ones. 

Let's start with the antisymmetric ones. These must vanish at the origin, $\hat f_n(x=0)=0$, which means that 
\eqn\ssss{{\rm Ai}\left(-ab_n\right)=0,\quad b_n = {\bar\gamma_n\over \pi}.
}
So, $-ab_n$ must be zeros of the Airy function ${\rm Ai}$. 

Similarly, for the symmetric wavefunctions, the derivative of the wavefunction must vanish at $x=0$. Therefore, for the symmetric ones, $-ab_n$ must be zeros of the derivative of the Airy function ${\rm Ai}'$. 

For highly excited states, the authors \ZubovURA\ assert that the values $\bar\gamma_n$ have the asymptotic behavior (see also \Cornwall)
\eqn\tttt{\bar\gamma_n\simeq \left[3\pi^2\left(n+\half\right)\right]^{2\over3}.
}
We show this by applying semiclassical quantization to classical periodic orbits later in section four. 

Another interesting question is, what is the momentum space wavefunction $f(s)$ \nnnn ? To compute it we need to do the inverse Fourier transform 
\eqn\uuuu{f(s)=\int_{-\infty}^\infty dx \hat f(x) e^{-isx}.
} 

We start by breaking the integral \uuuu\ into two parts, 
\eqn\fvvvv{f_n(s) = \int_{-\infty}^{\infty}{\hat f_n}(x)e^{-isx}dx = \int_{-\infty}^{0}{\hat f_n}(x)e^{-isx}dx  + \int_{0}^{\infty}{\hat f_n}(x)e^{-isx}dx,
}
where $n$ is a positive integer and labels the zeros $\bar\gamma_n$. For even $n$ since ${\hat f_n}$ is invariant under parity, we have
\eqn\fwwww{f_{2n}(s) =  2\int_{0}^{\infty}{\hat f_{2n}}(x)\cos(sx)dx,
}
and for odd $n$ ${\hat f}_n$ picks a minus sign under parity and thus, we have
\eqn\fxxxx{-if_{2n + 1}(s) =  -2\int_{0}^{\infty}{\hat f_{2n + 1}}(x)\sin(sx)dx.
}
The integral that we need to evaluate, therefore, using $\qqqq$ and $\rrrr$, is given by 
\eqn\fyyyy{I_{n}(s): = 2\int_{0}^{\infty}{\hat f_{n}}(x)e^{-isx}dx = 2\int_{0}^\infty dz{\rm Ai}(a z - a b_{n})e^{-isz}.
}
The real part of $I_{2n}$ gives \fwwww, and the imaginary part of $I_{2n + 1}$ gives \fxxxx.

The Airy function ${\rm Ai}(x)$ is an entire function with zeros located on the negative real axis. Therefore, it can be written as
\eqn\azzzz{{\rm Ai}(z) = \sum_{k = 0}^\infty c_k z^k,
}
where $c_k$ is a constant. Using this and performing a term by term integration we find\foot{In general we cannot exchange integration and sum unless the sum $I_n$ exists.}
\eqn\faaa{
I_n = -2ie^{-isb_n}\sum_{k = 0}^\infty c_k(ia)^k{d^k\over ds^k} {1\over s}\left(e^{isb_n} - \delta_{s,0}\right) = -2ie^{-isb_n}{\rm Ai}\left(ia{d\over ds}\right){1\over s}\left(e^{isb_n} - \delta_{s,0}\right).
}
We use this result shortly. See Fig. 1 for numerical plots of the momentum space wavefunction $f_n(s)$, \ie, $\fwwww$ and $\fxxxx$, for $n = 0, 1, 2, 3, 4$ and $5$.
 
We now estimate the order of the error that we earlier introduced in \mmmm\ while taking the interval of integration length from $1/t$ to infinity. To estimate the order of the error, therefore, we only need the asymptotic behavior of $f(s)$ for large $s$. 

In the large $s$ limit we have
\eqn\laaa{I_n = -{2i\over s}\sum_{m = 0}^{\infty} \left(ai\over s\right)^m {\rm Ai}^{(m)}(-ab_n) = -{2i{\rm Ai}(-ab_{n})\over s} + {2a{\rm Ai}'(-ab_{n})\over s^2}  + {\cal O}(s^{-3}).
}
where ${\rm Ai}^{(m)}$ is the $m$--th derivative of ${\rm Ai}$. Thus, the term that we ignored in \hhhh\ in taking the limits of integration to infinity, for odd $n$, is of order
\eqn\eaaa{\int_{{1\over t}}^{\infty}{dsf_{n}(s)\over s^2} \approx -ia^4{\rm Ai}^{(4)}(-ab_{n}) \int_{{1\over t}}^{\infty}{ds\over s^7} \approx -ia^4{\rm Ai}^{(4)}(-ab_{n}) t^{6}.
}
Similarly, for even $n$, the error is of order
\eqn\ebbb{\int_{{1\over t}}^{\infty}{dsf_{n}(s)\over s^2} \approx a^3{\rm Ai}^{(3)}(-ab_{n})\int_{{1\over t}}^{\infty}{ds\over s^6} \approx a^3{\rm Ai}^{(3)}(-ab_{n})t^{5}.
}
We note from \hhhh\ that the contribution from odd $n$ however cancels since the integrand is odd under $s \to -s$. Thus, the error we introduced by extending the integration limit to infinity comes only from even $n$ and it is of order $t^5$. This is in agreement with \Ziyatdinov.

Note that we also ignored the $\omega^2$ in the denominator of the first term on the r.h.s. of \hhhh\ which is of order $t^2$. Therefore, we are only considering the order $t$ correction. As a result, at order $t$ we can freely extend the limit of integration to infinity.

\bigskip

\ifig\loc{Numerical plots of the momentum space wavefunction $f_n(s)$. On the left hand side we have $f_n(s)$ for $n = 0$(orange), $n = 2$(black) and $n = 4$(purple). On the right hand side we have $if_n(s)$ for $n = 1$(orange), $n = 3$(black) and $n = 5$(purple). We note that the wavefunctions go to zero for large $s$.}
\centerline{\epsfxsize5.4in\epsffile{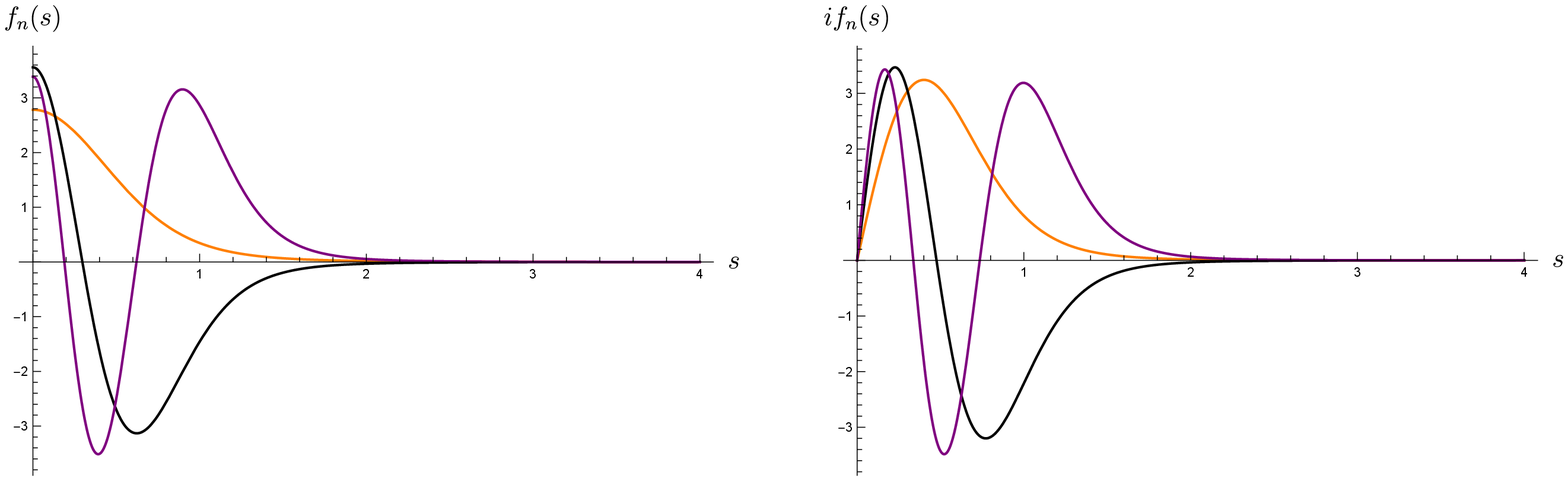}}

\bigskip

We next apply the above method to the two dimensional adjoint $QCD$.  

\newsec{The large mass limit of 2d adjoint $QCD$}

The theory is described by the action \refs{\Kutasov,\ \Dalley,\ \Bhanot},
\eqn\accc{S = \int d^2x \ \tr \left(i{\bar q}\gamma^{\alpha} D_{\alpha} q - m{\bar q}q - {1\over 4g^2}F_{\alpha\beta}F^{\alpha\beta} \right),
}
where the matrices $\gamma^0 = \eta^{00}\gamma_0 = \gamma_0, \gamma^{1} = \eta^{11}\gamma_1 = -\gamma_1, \{\gamma^{\alpha}, \gamma^{\beta}\} = 2\eta^{\alpha\beta}I_{2\times 2}$ are the $2 \times 2$ Dirac matrices in the Majorana representation, the field--strength tensor $F_{\alpha\beta} = \partial_{[\alpha}A_{\beta]} + iA_{[\alpha}A_{\beta]}$, the covariant derivative $D_{\alpha} = \partial_{\alpha} + i[A_{\alpha}, \cdot]$ and the fermion $q$ is a two component (Majorana--Weyl) spinor in the adjoint representation. We denote its top component as $\psi$ and bottom component as $\bar\psi$. The fermions $\psi$ and $\bar\psi$ are $N \times N$ hermitian traceless matrices. The gauge potential $A_{\alpha}$ is an $N \times N$ hermitian traceless matrix. $m$ is the bare fermion mass\foot{See \refs{\GrossM,\ \KutasovV} for a discussion on the massless case.}  and $g$ is the gauge coupling.\foot{Note that in two dimensions the gauge coupling $g$ is dimensionful.}

It is very convenient to use light--cone quantization \refs{\Hornbostel,\ \Pauli}. We introduce the light--cone coordinates by the definitions 
\eqn\laaa{x^{\pm} = {x^0 \pm x^1\over \sqrt{2}}.
}
We treat $x^+$ as the time variable. A useful gauge is $A_- = 0$. In this gauge we find 
\eqn\saaa{S = \int dx^+ dx^- \tr \left(i\psi \partial_+\psi + i\bar\psi\partial_-\bar\psi - i\sqrt{2}m\bar\psi\psi + {1\over 2g^2}(\partial_-A_+)^2 + A_+J^+\right),
}
where
\eqn\caaa{J^+_{ij} = 2\psi_{ik}\psi_{kj},
}
is an $SU(N)$ current. The gauge potential $A_+$ and the left moving fermion $\bar\psi$ are non--dynamical and can be eliminated using their equations of motion. We write the gauge potential $A_+ = A_{+, 0} + {\bar A}_+$, where $A_{+, 0}$ is the zero mode. Using the variational principle of least action we find
\eqn\ebbb{\int dx^-J^+ = 0, \quad \partial^2_-{\bar A}_+ - g^2J^+ = 0, \quad \sqrt{2}\partial_-{\bar\psi} - m\psi = 0.
}
Using these, the light--cone momentum and energy are given by 
\eqn\mbbb{
P^+ = \int dx^- \tr \left(i\psi\partial_-\psi\right)^2,
}
\eqn\eccc{
P^- = {1\over 2} \int dx^- \tr\left(im^2\psi{1\over \partial_-}\psi - g^2J^+{1\over \partial_-^2}J^+\right).
}
We now quantize the theory at $x^+ = 0$. We write the fermions as
\eqn\fddd{\psi_{ij}(x^-) = {1\over 2\sqrt{\pi}}\int_{-\infty}^{\infty} dk \psi_{ij}(k)e^{-ikx^-}.
}
The modes $\psi_{ij}(k)$ with $k < 0$ are creation operators and the modes $\psi_{ij}(k)$ with $k \geq 0$ are annihilation operators. 

The fermion modes satisfy the canonical anti--commutation relation given by
\eqn\aaa{\{\psi_{ab}(k),\psi_{cd}(k')\}=\delta(k+k')\left(\delta_{ad}\delta_{bc} - {1\over N}\delta_{ab}\delta_{cd}\right).
}
In terms of the modes, the translation generators in the large $N$ limit takes the form 
\eqn\ccc{\eqalign{P^+=&\int_0^\infty dk k\psi_{ab}(-k)\psi_{ba}(k),\cr
P^- = &\ {1\over 2} m^2 \int_0^\infty {dk\over k} \psi_{ab}(-k)\psi_{ba}(k)+
{1\over 2} g^2\int_0^\infty {dk\over k^2} J^+_{ab}(-k)J^+_{ba}(k),
}
}
where the current Fourier transform is given by
\eqn\jddd{J^+(k) = {1\over \sqrt{2\pi}}\int_{-\infty}^{\infty} dx^- J^+(x^-)e^{-ik x^-}.
}
Upon writing the current in terms of the modes, we find that, in the large $N$ limit, the light--cone Hamiltonian operator is given by
\eqn\peee{\eqalign{P^-
&= {1\over 2} m^2 \int_0^\infty {dk\over k} \psi_{ji}(-k)\psi_{ij}(k) + {g^2 N\over 2\pi}\int_0^\infty {dk\over k} C(k)\psi_{ji}(-k) \psi_{ij}(k)\cr
& - {g^2\over 2\pi}\int_0^\infty dk_1dk_2dk_3dk_4A(k_1, k_2, k_3, k_4)\delta(k_1 + k_2 - k_3 - k_4)\psi_{ij}(-k_4)\psi_{jk}(-k_3)\psi_{kl}(k_1)\psi_{li}(k_2) \cr
& + {g^2\over 2\pi}\int_0^\infty dk_1dk_2dk_3dk_4 B(k_1, k_2, k_3, k_4)\delta(k_1 + k_2 + k_3 - k_4)\cdot\cr
&\left[\psi_{jk}(-k_4)\psi_{kl}(k_1)\psi_{li}(k_2)\psi_{ij}(k_3) + \psi_{il}(-k_3)\psi_{lj}(-k_2)\psi_{jk}(-k_1)\psi_{ki}(k_4)\right],
}}
where
\eqn\defA{A(k_1, k_2, k_3, k_4) = {1\over(k_4 - k_2)^2} - {1\over (k_1 + k_2)^2},
}
\eqn\defB{B(k_1, k_2, k_3, k_4) = {1\over(k_2 + k_3)^2} - {1\over (k_1 + k_2)^2},
}
\eqn\defC{C(k) = \int_{0}^{\infty}dp\left[{k\over (p - k)^2} - {k\over (p + k)^2}\right]. 
}
We simplify $C(k)$ further as 
\eqn\defCP{\eqalign{C(k) 
& = \int_{0}^{\infty}dp\left[{k\over (p - k)^2} - {k\over (p + k)^2}\right],  \cr
& = \lim_{\epsilon\to 0}\left( \int_{0}^{k}dp{k\over (p - k - \epsilon)^2} + \int_{k}^{\infty}dp{k\over (p - k + \epsilon)^2}\right) - \int_0^\infty dp{k\over (p + k)^2}, \cr
& = 2\int_0^kdp{k\over (p - k)^2}.
}}

The light cone vacuum $|0\rangle$ is the ground state of $P^-$ with eigenvalue zero. All the physical states $|\chi\rangle$ must satisfy the zero charge constraint 
\eqn\crrr{\int dx^- J^+|\chi\rangle = 0.
}

The Hilbert space that the translation generators are taken to act on, in the large $N$ limit, is the space spanned by states of the form
\eqn\hbbb{\tr\left[\psi(-k_1)\psi(-k_2)\cdots \psi(-k_n)\right]|0\rangle.
}
These states satisfy the zero charge constraint. From the first line in \ccc\ we see that the total $P^+$ of a state of the form \hbbb, $k^+$, is 
\eqn\ddd{k^+=\sum_{i=1}^n k_i.
}
It is diagonal on the states \hbbb. To solve the theory, we need also to diagonalize the light--cone Hamiltonian, $P^-$ \ccc, on these states. In general, this is hard, since $P^-$ relates states with different values of the quark or parton number $n$ \hbbb. However, one may hope that this effect becomes less significant in the limit 
\eqn\eee{\lambda\equiv {g^2N\over m^2}\to 0.
}
This limit is the weak coupling limit of the theory. Indeed, one can think of $\lambda$ as the size of the (`t Hooft) coupling at the scale $m$, which is the scale associated with the bound states in this theory.\foot{And with the process of pair creation of the adjoint quarks.} 

Let's start with the free theory, \ie\ $\lambda=0$. In that case, $P^-$ \ccc\ is also diagonal on the states \hbbb, and we can compute its value, $k^-$,
\eqn\fff{k^-= {m^2\over 2}\sum_{i=1}^n {1\over k_i}.
}
It is useful to define the variables $x_i$ via 
\eqn\ggg{k_i=x_i k^+.
}
These variables take value in $(0,1)$ and can be thought of as the light--cone momentum fraction carried by the $i$'th parton. Obviously, one has (from \ddd)
\eqn\hhh{\sum_{i=1}^nx_i=1.
}
In terms of $x_i$, \fff\ can be written as 
\eqn\iii{M^2 = 2k^+k^-= m^2\sum_{i=1}^n{1\over x_i}.
}
The smallest value this quantity can take is $M=mn$, which is obtained by setting all $x_i$ to be equal to $1/n$. Moving away from this value, $M^2$ increases, and it diverges when any of the $x_i\to 0$. Thus, in the free theory, \ie \ $\lambda=0$, we find a continuum of masses starting at $mn$, precisely as we would expect for states of $n$ free particles. Now, we would like to turn on the leading effect of the interaction in \ccc. 

Consider, as an example, bound states consisting of two quarks. We can write these states in general as
\eqn\jjj{|\phi\rangle=\int_0^1 dx\phi(x)\tr\left[\psi(-xk^+)\psi(-(1-x)k^+)\right]|0\rangle,
}
where $\phi(x)$ is the wavefunction associated with the state. We saw earlier that for $\lambda=0$ the states that minimize the energy correspond to wavefunctions that are very sharply peaked around $x=1/2$. Such states have mass $M\sim 2m$, the mass of a state of two free quarks. In general, we will choose the wavefunction $\phi(x)$ to satisfy the boundary condition 
\eqn\kkk{\phi(0)=0,
}
this is consistent with our definition of the modes. Note that the wavefunction is by definition antisymmetric under $x\to 1-x$, 
\eqn\lll{\phi(1-x)=-\phi(x).
}
Thus, \kkk\ also implies vanishing of the wavefunction at $x=1$. The inner product between two states of the form \jjj, $|\phi\rangle$ and $|\phi'\rangle$, is given by 
\eqn\mmm{\langle\phi'|\phi\rangle={2N^2\over k^+}\delta(k^+-{k^{+}}')\int_0^1dx\phi(x)\phi'(x).
}
In particular, the norm $\langle\phi|\phi\rangle$, is positive definite, as expected.

Similarly, we can define a general $n$ partons gauge invariant bound state as 
\eqn\state{\left|\phi\right\rangle := \int_0^{k^+} dk_1\cdots dk_n \delta\left(\sum_{i = 1}^nk_i - k^+\right)\phi_n(k_1, \cdots, k_n){\rm tr}\left[\psi(-k_1)\cdots \psi(-k_n)\right]\left|0\right\rangle,
}
where $\phi_n$ is the wavefunction associated with the state $\left|\phi\right\rangle$. Therefore, for even number of partons the state is bosonic and for odd number of partons the state is fermionic. Note that by definition the wavefunction has the property
\eqn\saaaaa{ \phi_n(k_1, k_2, \cdots, k_{n - 1}, k_n) = (-1)^{n - 1}\phi_n(k_2, k_3, \cdots, k_n, k_1).
} 
We will choose the wavefuncton, in general, to satisfy the condition
\eqn\caaaa{ \phi_n(0, k_2, \cdots, k_{n - 1}, k_n) = 0,
}
this is consistent with our definition of the modes.

Acting with the light cone Hamiltonian $P^-$ \peee\ on the state $\left|\phi\right\rangle$ gives the following equation for the $M^2_n$ of the state.
\eqn\statePP{\eqalign{M^2_n\phi_n(x_1, \cdots, x_n)  
&= m^2\sum_{i = 1}^n{1\over x_i}\phi_n(x_1, \cdots, x_n)  \cr
& +{2g^2N\over\pi}\sum_{i = 1}^n \phi_n(x_1, \cdots, x_{i - 1}, x_{i }, x_{i + 1}, \cdots, x_n)\cdot  \int_0^{x_{i} }dy{1\over (y - x_i)^2}\cr
&+{g^2N\over \pi} \sum_{i = 1}^n {1\over \left(x_i + x_{i + 1}\right)^2}\int_0^{x_i + x_{i + 1}} dy \phi_n(x_1, \cdots, x_{i - 1}, y, x_i + x_{i + 1} - y, x_{i + 2}, \cdots ,x_n)\cr
& -{g^2N\over\pi}\sum_{i = 1}^n \int_0^{x_{i} + x_{i + 1}}dy\phi_n(x_1, \cdots, x_{i - 1}, y, x_{i }+ x_{i + 1} - y , x_{i + 2}, \cdots, x_n)\cdot {1\over (y - x_{i})^2}\cr
&+{g^2N\over \pi}\sum_{i = 1}^{n}\int_0^{x_i}dy\int_{0}^{x_i - y}dz \phi_{n + 2}(x_1, \cdots , x_{i - 1}, y, z, x_i - y - z, x_{i + 1}, \cdots, x_n)\cdot\cr
&\left[{1\over (y + z)^2} - {1\over (x_i - y)^2}\right]\cr
&+{g^2N\over\pi}\sum_{i = 1}^{n} \phi_{n - 2}(x_1, \cdots, x_{i - 1}, x_i + x_{i + 1} + x_{i + 2}, x_{i + 3}, \cdots, x_n)\cdot\cr
& \left[{1\over (x_i + x_{i + 1})^2} - {1\over (x_{i + 1} + x_{i + 2})^2}\right], \quad x_1 = x_{n + 1}, \quad \sum_{i = 1}^n x_i = 1.
}}

We note that for even values of $n$ the equation only involves bosonic states, and similarly, for odd values of $n$ it only involves fermionic states. Thus, it does not mix bosonic and fermionic states. We also note that the equation relates or mixes states with different partons number $n, n \pm 2$. This is the main reason why solving this equation analytically and exactly, even in the planar limit, has been difficult. We rewrite this equation using the redefinitions
\eqn\defM{\pi M^2_n = g^2N\mu^2_n, \quad \alpha = {m^2\pi\over g^2 N}, 
}
as
\eqn\stateP{\eqalign{\mu^2_n\phi_n(x_1, \cdots, x_n)  
&= \alpha\sum_{i = 1}^n{1\over x_i}\phi_n(x_1, \cdots, x_n)\cr
&+\sum_{i = 1}^{n}\int_0^{x_i}dy\int_{0}^{x_i - y}dz \phi_{n + 2}(x_1, \cdots , x_{i - 1}, y, z, x_i - y - z, x_{i + 1}, \cdots, x_n)\cr
 &\left[{1\over (y + z)^2} - {1\over (x_i - y)^2}\right]\cr
&+ \sum_{i = 1}^n {1\over \left(x_i + x_{i + 1}\right)^2}\int_0^{x_i + x_{i + 1}} dy \phi_n(x_1, \cdots, x_{i - 1}, y, x_i + x_{i + 1} - y, x_{i + 2}, \cdots ,x_n)\cr
 & +\sum_{i = 1}^n \int_0^{x_{i} + x_{i + 1}} {dy\over (y - x_{i})^2} \cr
 &\left[\phi_n(x_1, \cdots, x_{i - 1}, x_{i }, x_{i + 1}, \cdots, x_n) - \phi_n(x_1, \cdots, x_{i - 1}, y, x_{i }+ x_{i + 1} - y , x_{i + 2}, \cdots, x_n)\right]\cr
&+\sum_{i = 1}^{n} \phi_{n - 2}(x_1, \cdots, x_{i - 1}, x_i + x_{i + 1} + x_{i + 2}, x_{i + 3}, \cdots, x_n)\cr
& \left[{1\over (x_i + x_{i + 1})^2} - {1\over (x_{i + 1} + x_{i + 2})^2}\right], \quad x_1 = x_{n + 1}, \quad \sum_{i = 1}^n x_i = 1.
}}
Here we have used the identity
\eqn\defiii{\int_{x_i}^{x_i + x_{i + 1}}dy {1\over{(y - x_i)^2}} = \int_{0}^{x_{i + 1}}dy{1\over(y - x_{i + 1})^2}.
}

We now write as we did in the previous section 
\eqn\aroundMI{x_i = {1\over n} + \omega_i, \quad \sum_{i = 1}^n \omega_i = 0.
}
In the following analysis we will assume $n\omega_i \ll 1$. Therefore, the states are sharply picked around $x_1 = x_2 = \cdots = x_n = 1/n$. We also define 
\eqn\muU{\mu^2_n = n^2\alpha + \gamma_n.
}
We are interested in the large $\alpha$ limit such that
\eqn\limiT{\lim_{\alpha\to 0}{\gamma_n\over n^2\alpha} \to 0.
}
Using the above redefinitions, the l.h.s. of \stateP\ becomes
\eqn\rightHM{n^2\alpha\phi_n(x_1, \cdots, x_n) + \gamma_n\phi_n(x_1, \cdots, x_n).
}
We next look the r.h.s. of the equation \stateP\ term by term. 

From the first term we have
\eqn\firsT{\alpha\sum_{i = 1}^n{1\over x_i}\phi_n(x_1, \cdots, x_n) = \alpha\sum_{i = 1}^n{n\over 1 + n\omega_i}\phi_n(x_1, \cdots, x_n) = n^2\alpha\phi_n + n^3\alpha\sum_{i = 1}^n\omega_i^2\phi_n+ {\cal O}(n^4\alpha\omega_i^3)\phi_n.
}

From the second term with
\eqn\defYZ{y = {1\over n} + \omega_y, \quad z = {1\over n} + \omega_z, \quad n\omega_y \ll 1, \quad n\omega_z \ll 1,
}
we have
\eqn\statePS{\eqalign{\sum_{i = 1}^{n}\int_0^{x_i}dy\int_{0}^{x_i - y}dz \phi_{n + 2}
 \left[{1\over (y + z)^2} - {1\over (x_i - y)^2}\right]\cr
=  \sum_{i = 1}^{n}\int_{-{1\over n}}^{\omega_i}d\omega_y\int_{-{1\over n}}^{-{1\over n} + \omega_i - \omega_y}d\omega_z \phi_{n + 2}
 \left[{n^2\over 4} - {1\over (\omega_i - \omega_y)^2} + {\cal O}(n\omega_y + n\omega_z)\right]. &
}}

From the third term we find
\eqn\statePTH{\eqalign{\sum_{i = 1}^n {1\over \left(x_i + x_{i + 1}\right)^2}\int_0^{x_i + x_{i + 1}} dy \phi_n = \sum_{i = 1}^n {1\over \left({2\over n} + \omega_i + \omega_{i + 1}\right)^2}\int_{-{1\over n}}^{{1\over n} + \omega_i + \omega_{i + 1}} d\omega_y \phi_n, \cr
 =   \sum_{i = 1}^n\left( {n^2\over 4} - {n^3\over 4}(\omega_i + \omega_{i + 1})+ {\cal O}(\omega_i^2)\right)\int_{-{1\over n}}^{{1\over n} + \omega_i + \omega_{i + 1}} d\omega_y \phi_n.&
}}

From the fourth term we get
\eqn\statePFR{\eqalign{\sum_{i = 1}^n \int_0^{x_{i} + x_{i + 1}}{dy\over (y - x_{i})^2}\cdot\left[\phi_n(x_1, \cdots, x_n) - \phi_n(x_1, \cdots, y, x_i + x_{i + 1} - y, \cdots, x_n)\right] \cr
=\sum_{i = 1}^n \int_{-{1\over n}}^{{1\over n} + \omega_i + \omega_{i + 1}} {d\omega_y\over (\omega_y - \omega_{i})^2}\cdot\left[\phi_n(\omega_1, \cdots, \omega_n) - \phi_n(\omega_1, \cdots, \omega_y, \omega_i + \omega_{i + 1} - \omega_y, \cdots, \omega_n)\right]. &
}}

From the last term we get
\eqn\statPLA{\eqalign{\sum_{i = 1}^{n} \phi_{n - 2} \left[{1\over (x_i + x_{i + 1})^2} - {1\over (x_{i + 1} + x_{i + 2})^2}\right] = \sum_{i = 1}^{n} \phi_{n - 2} \left[{1\over ({2\over n} + \omega_i + \omega_{i + 1})^2} - {1\over ({2\over n} + \omega_{i + 1} + \omega_{i + 2})^2}\right], \cr
= \sum_{i = 1}^{n} \phi_{n - 2} \cdot {n^2\over 4}\left[ n\left(\omega_{i + 2} - \omega_i\right) - {3n^2\over 4}\left(\omega_{i + 2} - \omega_i\right)\left(\omega_i + 2\omega_{i + 1} + \omega_{i + 2}\right) +  {\cal O}(\omega_i^3)\right].  &
}}

We next rescale the $\omega_i$'s as 
\eqn\defOM{\omega_i = s_i t, \quad \omega_y = s_y t, \quad \omega_z = s_z t.
}
As we did in the previous section, we assume that the wavefunctions 
\eqn\indeP{\phi_n(\omega_1, \cdots, \omega_i, \cdots, \omega_n) := \phi_n(s_1, \cdots, s_i, \cdots, s_n),
}
do not depend on $t$. That is, the wavefunctions are sharply picked around $x_1 = x_2 = \cdots = x_n = 1/n$. We take 
\eqn\defTT{t = \alpha^{-{1\over 3}},
}
and redefine $\gamma_n$ as
\eqn\redfGA{\bar\gamma_n = t\gamma_n.
}
In the large $\alpha$ limit we then get
\eqn\massSQ{\eqalign{\bar\gamma_n \phi_n 
&=n^3\sum_{i = 1}^n s_i^2\phi_n - tn^4\sum_{i = 1}^ns_i^3\phi_n +  {\cal O}(t^2)\phi_n\cr
&+t\sum_{i = 1}^{n}\int_{-{1\over nt}}^{s_i}ds_y\int_{-{1\over nt}}^{-{1\over nt} + s_i - s_y}ds_z \phi_{n + 2}
 \left[{n^2t^2\over 4} - {1\over (s_i - s_y)^2} + {\cal O}(t^3)\right] \cr
&+t^2\sum_{i = 1}^n\left( {n^2\over 4} - {n^3t\over 4}(s_i + s_{i + 1}) + {\cal O}(t^2)\right)\int_{-{1\over nt}}^{{1\over nt} + s_i + s_{i + 1}} ds_y \phi_n\cr
 & + \sum_{i = 1}^n\int_{-{1\over nt}}^{{1\over nt} + s_i + s_{i + 1}}{ds_y\over (s_y - s_i)^2} \cdot\left[\phi_n(s_1, \cdots, s_n) - \phi_n(s_1, \cdots, s_y, s_i + s_{i + 1} - s_y, \cdots, s_n)\right]  \cr
&+t^2\sum_{i = 1}^{n} \phi_{n - 2} \cdot {n^2\over 4}\left[ n\left(s_{i + 2} - s_i\right) -{3n^2t\over 4}\left(s_{i + 2} - s_i\right)\left(s_i + 2s_{i + 1} + s_{i + 2}\right) +  {\cal O}(t^2)\right].
}}

Therefore, to leading order, we have the mass squared equation
\eqn\massSQG{\eqalign{\bar\gamma_n \phi_n 
&=n^3\sum_{i = 1}^n s_i^2\phi_n \cr
& + \sum_{i = 1}^n\int_{-\infty}^{\infty}{ds_y\over (s_y - s_i)^2} \cdot\left[\phi_n(s_1, \cdots, s_n) - \phi_n(s_1, \cdots, s_y, s_i + s_{i + 1} - s_y, \cdots, s_n)\right]  \cr
&- tn^4\sum_{i = 1}^ns_i^3\phi_n +  {\cal O}(t^2).
}}

Note that at this order, \ie\ ${\cal O}(t)$, only $\phi_n$ contributes to the mass squared equation. Thus, for the low--lying states, there is no pair production or annihilation, as expected. This was noted already in \refs{\Dalley,\ \Bhanot}, and there is also recent numerical evidence that suggests this is the case for the low--lying states even at moderate values of the coupling \DempseyY. Note also that, at this order, we see using \stateP\ that \massSQG\ is equivalent to the $n$--parton `t Hooft equation  
\eqn\stateDKP{\eqalign{\mu^2_n\phi_n(x_1, \cdots, x_n) 
& = \alpha\sum_{i = 1}^n{1\over x_i}\phi_n(x_1, \cdots, x_n) + \sum_{i = 1}^n \int_0^{x_{i} + x_{i + 1}} {dy\over (y - x_{i})^2}\cdot \cr
 &\left[\phi_n(x_1, \cdots, x_{i - 1}, x_{i }, x_{i + 1}, \cdots, x_n) - \phi_n(x_1, \cdots, x_{i - 1}, y, x_{i }+ x_{i + 1} - y , x_{i + 2}, \cdots, x_n)\right].
}}
Therefore, the goal is to solve this equation in the region in which the momentum fractions $x_1 = \cdots = x_n$ are near $1/n$. In particular, for $n = 2$, we have,
\eqn\stateDKPP{\mu^2_2\phi_2(x) = {\alpha\over x(1 - x)}\phi_2(x) - 2\int_{0}^{1}{dy\over (y - x)^2}\phi_2(y).
}
Note that the integral is defined in the principal value sense, see \defCP. This is the `t Hooft equation \aaaa.\foot{See also \Dhar\ for a similar equation obtained using a formulation of 2d fundamental QCD in terms of bilocal fields and the method of coadjoint orbits.} The source of the extra factor 2 will be discussed shortly.

We write the Fourier transform of the wavefunction $\phi_n$ as
\eqn\FourierNN{\hat\phi_n(\vec{x}) := {1\over (2\pi)^n}\int  e^{i\vec{ x}\cdot \vec{s}}\cdot \delta(s_1+\cdots+s_n)\phi_n(\vec{ s})d{\vec{ s}},
}
equivalently 
\eqn\FourierNM{\delta(s_1+\cdots+s_n)\phi_n(\vec{s}) = \int  e^{-i\vec{ x}\cdot \vec{s}}\hat\phi_n(\vec{ x})d{\vec{ x}}.
}

To do the Fourier transform of  the mass squared equation \massSQG\ we need the value of the integral 
\eqn\intGG{\int_{-\infty}^{\infty} {{e^{-i({ x}_i - { x}_{i + 1})s_y}}\over (s_y - s_i)^2}ds_y.
}
As in the t' Hooft model, the integral is defined by a principal value prescription. We assume the following integration prescription\foot{This is similar to \defCP.}  
\eqn\princPAL{P\int {f(s)\over (s - s_0)^2}ds = {1\over 2}\int {f(s)\over (s - s_0 + i\epsilon)^2}ds + {1\over 2}\int {f(s)\over (s - s_0 - i\epsilon)^2}ds.
}
Using this prescription we get
\eqn\intGGG{\int_{-\infty}^{\infty} {{e^{-i({ x}_i - { x}_{i + 1})s_y}}\over (s_y - s_i)^2}ds_y = -\pi|{ x}_i - { x}_{i + 1}|e^{-i({ x}_i - { x}_{i + 1})s_i}.
}
Using the above result we see that
\eqn\simPP{\eqalign{\int_{-\infty}^{\infty}{ds_y\over (s_y - s_i)^2}\delta(s_1 +\cdots + s_n)\phi_n(s_1, \cdots, s_y, s_i + s_{i + 1} - s_y, \cdots, s_n)
&\cr
= -\pi\int |y_i - y_{i + 1}|e^{-i\vec{y}\cdot \vec{s}}\hat\phi_n(\vec{y})d\vec{y}.
}}
Therefore, to order ${\cal O}(t)$, the Fourier transform of equation \massSQG\ becomes
\eqn\FouriRR{\bar\gamma_n\hat\phi_n(\vec{x}) = -n^3\sum_{i = 1}^n \partial_{x_i}^2\hat\phi_n({\vec{x}}) + \pi\sum_{i = 1}^n|x_i - x_{i + 1}|\hat\phi_{n}(\vec{x}) + {\cal O}(t).
} 

The wavefunction $\hat{\phi}_n$ has the following symmetries
\eqn\syMM{\hat\phi_n(x_1 + c, \cdots, x_n + c) = \hat{\phi}_n(x_1, \cdots, x_n), \quad \hat{\phi}_n(x_1, x_2, \cdots, x_{n - 1}, x_n) = (-1)^{n - 1}\hat{\phi}_n(x_2, x_3, \cdots, x_n, x_1),
} 
here $c$ is a constant.

We note that the $n$--parton bound state potential is given by a pairwise sum of two--parton potentials. This can be also seen directly from the $n$--parton `t Hooft equation \stateDKP. The doubling of the strength of the coulomb interaction or potential for $n = 2$ is due to the two color flux tubes connecting a pair of partons (in a quark anti--quark pair there is only one flux tube) (see, for example, \Bhanot). For three adjoint quarks the potential $V$ is given by
\eqn\THREEQV{V(x_1, x_2, x_3) = |x_1 - x_2| + |x_2 - x_3| + |x_3 - x_1| + {\cal O}(t).
}
In $1 + 1$d fundamental $QCD$ similar expression was obtained in \Hornbostell\ for a baryon, which is a bound state of three quarks, in the heavy--quark limit, see section (3.9) of the paper.\foot{I thank Igor Klebanov for bringing to my attention this result.} In ($1 + 3$d) $QCD$ there are two ans\"{a}tzes regarding the three quraks potential. They are known as the $\Delta$ and $Y$ ans\"{a}tzes. In the $\Delta$ ans\"{a}tz the potential is given by \THREEQV. There is no clear answer however regarding the correct three quarks static potential. For recent discussions on three quarks potential in phenomenological models of $QCD$ see \refs{\Andreev,\ \Bali}.

In the next section we discuss the cases $n = 2$ and $n = 3$. These cases can be easily generalized to the $n \geq 4$ cases in a similar manner. 

\newsec{Discussion}

We now discuss the $n$--body non--relativistic Schr\"{o}dinger equation 
\eqn\FouriRRM{\bar\gamma_n\hat\phi_n(\vec{x}) = -n^3\sum_{i = 1}^n \partial_{x_i}^2\hat\phi_n({\vec{x}}) + \pi\sum_{i = 1}^n|x_i - x_{i + 1}|\hat\phi_{n}(\vec{x}), \quad x_{n + 1} = x_1, 
} 
with the (boundary) conditions or constraints  
\eqn\syMM{\hat\phi_n(x_1 + c, \cdots, x_n + c) = \hat{\phi}_n(x_1, \cdots, x_n), \quad \hat{\phi}_n(x_1, x_2, \cdots, x_{n - 1}, x_n) = (-1)^{n - 1}\hat{\phi}_n(x_2, x_3, \cdots, x_n, x_1),
} 
for the cases where the partons number $n$ is $2$ and $3$. We begin our discussion with $n = 2$.

The $n = 2$ case is very similar to the (fundamental) `t Hooft model. In this case the Schr\"{o}dinger equation is 
\eqn\TWOAQ{\bar\gamma_2\hat\phi_2(x_1, x_2) = -16(\partial_{x_1}^2 + \partial_{x_2}^2)\hat\phi_2(x_1, x_2) + 2 \pi |x_1 - x_2| \hat\phi_{2}(x_1, x_2),  
}
and
\eqn\CONDTQ{\hat\phi_{2}(x_1, x_2) = -\hat\phi_{2}(x_2, x_1). 
}
It is very convenient to introduce the Jacobi coordinates 
\eqn\JJC{z_1 = x_1 - x_2, \quad z_2 = {x_1 + x_2\over 2}.
} 
In terms of which the equation becomes
\eqn\FouriRRJ{\bar\gamma_2\hat\phi_2 = -8\left(2\partial_{z_1}^2 + {1\over 2}\partial^2_{z_2}\right)\hat\phi_2 + 2\pi|z_1|\hat\phi_2.
} 
Since we are interested on bound states we set the center of mass coordinate $z_2$, using the translation symmetry, to zero. Therefore, the relative motion of the quarks is described by 
\eqn\FouriRRJ{\bar\gamma_2\hat\phi_2 = -16\partial_{z_1}^2 \hat\phi_2 + 2\pi|z_1|\hat\phi_2, \quad \hat\phi_2(z_1) = -\hat\phi_2(-z_1).
} 

After rescaling the coordinates, this can be put into the more familiar form 
\eqn\FouriRRJ{\gamma\phi = -{1\over 2}{d^2 \phi\over dz^2} + |z|\phi, \quad \gamma = {\bar\gamma_2\over 2\pi}\left(\pi\over 2\cdot 8\right)^{1\over 3},\quad \phi(z) = -\phi(-z).
} 
This is the Airy equation and its solutions are discussed in detail in section two. The wavefunction $\hat\phi_2$ in the adjoint case is given by
\eqn\TWOQS{\hat\phi^{(l)}_2(z) = \left\{\eqalign{
 {\rm Ai}\left(\left({\pi\over 8}\right)^{1\over 3}\left(z - {\bar\gamma_2^{(l)}\over 2\pi}\right)\right), &\quad  z > 0,\cr
 -{\rm Ai}\left(\left({\pi\over 8}\right)^{1\over 3}\left(-z - {\bar\gamma_2^{(l)}\over 2\pi}\right)\right),&\quad  z< 0,
}\right.
}
where $\bar\gamma_2^{(l)}$ are given by the equations 
\eqn\ZZERO{
{\rm Ai}\left(-\bar\gamma_2^{(l)}/2\pi(8\pi^2)^{1/3}\right) = 0,  \quad l = 1, 3, 5, \cdots.
}
Therefore, the masses are given by
\eqn\MASQQ{M_{(2, l)}^2 = m^2\left(4 + \lambda^{2\over 3} \bar\gamma_2^{(l)}\right), \quad \lambda := {g^2 N\over m^2\pi }, \quad l = 1, 3, 5, \cdots.
}

The quantum spectrum for the highly excited bound states can be computed by considering the periodic orbits of the corresponding classical Hamiltonian. The classical Hamiltonian in this case is
\eqn\cllH{H = {p^2\over 2} + |z|.
}
A typical periodic motion in this system is described by 
\eqn\eqmCSE{z(t) = \left\{\eqalign{
 & -t(t - t_2), \quad 0 \leq t \leq t_2,\cr
 & (t - t_2)(t - 2t_2), \quad t_2 \leq t \leq 2t_2,
}\right.
}
here $T = 2t_2$ is the period. We now apply the Einstein--Brillouin--Keller (EBK) quantization. We first evaluate the action integral 
\eqn\ebK{2\int_0^{T\over 2}p^2 dt = {T^3\over 24} = {8\over 3}E^{3\over 2},
}
where $E$ is the energy of the system along the orbit. This gives making use of the EBK quantization condition the spectrum 
\eqn\statE{E_n = \left[{3\over 4}\pi\left(n + {1\over 2}\right) \right]^{2\over 3}.
}
From this it follows that 
\eqn\gaa{\bar\gamma_2^{(n)} = 2\pi \left(2\cdot 8\over \pi \right)^{1\over 3} E_n = 2\left[3\pi^2\left(n + {1\over 2}\right) \right]^{2\over 3}.
}
Note the factor of 2 due to the two flux tubes. In one dimension EBK is similar to Wentzel--Kramers--Brillouin (WKB) approximation. Putting all together we have for the highly excited states 
\eqn\MSS{M_{(2, n)}^2 = m^2\left\{4 + 2\cdot \lambda^{2/3}\left[{3\pi^2 }\left(n + {1\over 2}\right) \right]^{2\over 3}\right\}, 
}
where $n$ is odd and large integer and $\lambda$ is the `t Hooft coupling \MASQQ\ at the scale of the constituent quark mass $m$.

We now consider the three partons case. In this case the Schr\"{o}dinger equation takes the form
\eqn\FouriRRTH{\bar\gamma_3\hat\phi_3(\vec{x}) = -27(\partial_{x_1}^2 + \partial_{x_2}^2 + \partial_{x_3}^2)\hat\phi_3({\vec{x}}) + \pi(|x_1 - x_2| + |x_2 - x_3| + |x_3 - x_1|)\hat\phi_{3}(\vec{x}),  
} 
with the constrains on the wavefunction 
\eqn\syMM{\hat\phi_3(x_1 + c, \cdots, x_3 + c) = \hat{\phi}_3(x_1, \cdots, x_3), \quad \hat{\phi}_n(x_1, x_2, x_3) = \hat{\phi}_3(x_2, x_3, x_1) = \hat{\phi}_3(x_3, x_1, x_2).
}
We next conjecture that this equation is solvable. In particular, after making a change of coordinates, we conjecture that it can be solved using the method of separation of variables. It is important that one makes a change to parabolic coordinates to solve the problem.\foot{I would like to mention that a similar equation to \FouriRRTH\ was previously obtained in \Hornbostell\ by K. Hornbostel for a baryon. I thank Igor Klebanov for brining this result to my attention. However, the equation was not solved. The author is not aware of any other work.} We expect that this generalizes to $n$--parton state.

We write \FouriRRTH\ as
\eqn\FouriRRTHHY{\gamma \psi = -{1\over 2}\left(\partial_{x_1}^2 + \partial_{x_2}^2 + \partial_{x_3}^2\right)\psi + (|x_1 - x_2| + |x_2 - x_3| + |x_3 - x_1|)\psi, \quad \gamma = {\bar\gamma_3\over\pi}\left(\pi\over 2\cdot 27\right)^{1\over 3}.
}
Since we are interested in the relative motion of the quarks we introduce the Jacobi coordinates
\eqn\JJC{z_1 = {x_1 + x_2 + x_3\over 3}, \quad z_2 = {x_2 - x_1\over \sqrt{2}}, \quad z_3 = \sqrt{2\over 3}\left(x_3 - {x_1 + x_2\over 2}\right).
}
We note that
\eqn\pot{x_1 - x_2 = -\sqrt{2}z_2, \quad x_2 - x_3 = {1\over \sqrt{2}}\left(z_2 - \sqrt{3}z_3\right), \quad x_3 - x_1 = {1\over \sqrt{2}}\left(z_2 + \sqrt{3}z_3\right),
}
The relative motion of the quarks then becomes 
\eqn\TTTQQ{\gamma \psi = -{1\over 2}\left(\partial_{z_2}^2 + \partial_{z_3}^2\right)\psi + V( z_2, z_3)\psi,
}
where the potential $V$ is given by
\eqn\VVTH{V( z_2, z_3) = \sqrt{2}|z_2| + {1\over\sqrt{2}}|z_2 - \sqrt{3}z_3| + {1\over\sqrt{2}}|z_2 + \sqrt{3}z_3|.
}

The equation can be written in a more familiar and useful form using the polar coordinates. We define
\eqn\redCY{z_2 = -r\sin\phi, \quad z_3 = -r\cos\phi,
}
where
\eqn\POO{ 0\leq r < \infty, \quad 0 \leq \phi < 2\pi.
}

The light--like separations of the partons are given in terms of the polar coordinates by
\eqn\POLZZ{
\eqalign{
 x_1 - x_2  = -\sqrt{2}z_2 & = \sqrt{2}r\sin\phi,\cr
 x_2 - x_3 = {1\over \sqrt{2}}\left(z_2 - \sqrt{3}z_3\right)& = \sqrt{2}r\sin\left(\phi + {2\over 3}\pi\right),\cr
 x_3 - x_1 = {1\over \sqrt{2}}\left(z_2 + \sqrt{3}z_3\right)& = \sqrt{2}r\sin\left(\phi + {4\over 3}\pi\right).
}
}
This follows from \redCY\ and \pot. The different sectors or orderings in the original and new coordinates are related as follows.
\eqn\eqmCSE{\eqalign{
x_1 > x_2 > x_3 &,\ \ie,\ 0 < \phi < {1\over 3}\pi,\cr
x_1 > x_3 > x_2 &,\ \ie,\ {1\over 3}\pi < \phi < {2\over 3}\pi,\cr
x_3 > x_1 > x_2 &,\ \ie,\ {2\over 3}\pi < \phi < \pi,\cr
x_3 > x_2 > x_1 &,\ \ie,\ \pi < \phi < {4\over 3}\pi, \cr
x_2 > x_3 > x_1 &,\ \ie,\ {4\over 3}\pi < \phi < {5\over 3}\pi,\cr
x_2 > x_1 > x_3 &,\ \ie,\  {5\over 3}\pi < \phi < 2\pi.
}
}

In terms of the polar coordinates the Schr\"{o}dinger equation now becomes
\eqn\soo{H\psi = \gamma\psi, 
}
where $H$ is the Hamiltonian 
\eqn\three{H = -{1\over 2}\left({\partial^2\over \partial r^2} + {1\over r}{\partial \over \partial r} + {1\over r^2}{\partial^2 \over \partial \phi^2}\right)  +  \sqrt{2} r\left(|\sin\phi| + |\sin\left(\phi + {2\pi\over 3}\right)| + |\sin\left(\phi + {4\pi\over 3}\right)|\right),
}
and the wavefunction satisfy the symmetry\foot{Recall that the fermions are hermitian. Also, in two dimensions we can impose simultaneously both Weyl and Majorana conditions.}
\eqn\bondD{  \psi\left(r, \phi \right) = -\psi\left(r, \phi + {\pi\over 3}\right), 
}
which also implies
\eqn\bond{\psi\left(r, \phi \right) = \psi\left(r, \phi + 2\pi \right).
}
Note also that the equation is invariant under the parity $\phi \to -\phi$. Thus, we only need to consider the sector
\eqn\threeMA{\left({1\over 2}p_r^2 + {1\over 2r^2}p_{\phi}^2 +  2\sqrt{2} r\sin\phi - \gamma\right)\psi(r, \phi) = 0, \quad {\pi\over 3} < \phi < {2\pi\over 3},
} 
with the anti--periodic boundary condition
\eqn\bondWQ{  \psi\left(r, {\pi\over 3} \right) = -\psi\left(r,  {2\pi\over 3}\right), 
}
and the usual boundary conditions at $r = 0$ (\ie, the wavefunction must be finite at the origin and it must be also single valued as we approach the origin from different angular directions) and $r = \infty$ (\ie, the wavefunction must be normalizable),
\eqn\bondWQR{ \psi\left(0,  \phi\right) = 0, \quad \psi\left(\infty,  \phi\right) = 0, 
}
where 
\eqn\threeev{p_r^2 = {\bar p}_r^2 - {1\over 4r^2},\quad { \bar p}_r = {-i\over r^{1\over 2}}{\partial\over \partial r}r^{1\over 2}, \quad p_{\phi} = -i{\partial\over \partial\phi},
} 
are the (generalized) radial and angular momenta operators. 

We slightly rewrite \threeMA\ and \bondWQ\ in the following form.
\eqn\threeRT{\left({1\over 2}p_r^2 + {1\over 2r^2}p_{\phi}^2 + p r\cos(\phi -\phi_0) - \gamma\right)\psi(r, \phi) = 0,  \quad \phi_0 - {\pi\over 6} < \phi < \phi_0 + {\pi\over 6},
}
where
\eqn\somDEF{ \psi(r, \phi_0 - \pi/6) = - \psi(r, \phi_0 + \pi/6),
} 
and the constant $p = 2\sqrt{2}$. The phase $\phi_0$ takes different values depending on which sector or domain $\phi$ belongs to. It only takes the values $\pm\pi/6, \pm\pi/2, \pm 5\pi/6$ since there are only six domains in total. In \threeMA\ $\phi_0 = \pi/2$. Also note that \threeRT\ and \somDEF\ are invariant under $\phi \to 2\phi_0 - \phi$. Thus, the equation has a $\Z_2$ reflection symmetry. 

We note that this equation together with the cyclic constraint \bondD\ describes a particle confined in (or to the surface of) an inverted hexagonal pyramid potential. The hexagonal base is at infinity. The potential has the same symmetry group as the base. In general, the general equation \FouriRRM\ describes a particle confined to the surface of an inverted $t$--gonal pyramid potential in $n$ dimensions. We saw that $t = 1$ for $n = 2$, and $t = n!$ for $n = 3.$

We also note that the equation \threeRT\ describes a non--relativistic point electric dipole in a plane angular sector with electric field proportional to $r$ in an appropriate unit.\foot{Alternatively it describes a unit point charge in a constant electric field background. The unit charge can be treated as a dipole with dipole moment proportional to ${\vec r}$. The constant electric field has the angular direction $\phi_0$. } The plane angular sector has (wedge) angle $\pi/3$. The quantity $p$ is the magnitude of the electric dipole moment in an appropriate unit and the phase $\phi_0$ is the electric field angular direction. In general, we expect such interpretation to arise in the general case \FouriRRM.

The trajectories of charged particles in the presence of electric field are parabolic. Thus, it is very convenient to use parabolic coordinates to simplify the equation further. We introduce the parabolic coordinates with the definitions 
\eqn\parab{r\sin\left(\phi - \phi_0\right) = \xi\sigma\tau, \quad r\cos\left(\phi - \phi_0\right) = {1\over 2}(\tau^2 - \sigma^2),
}
where $\xi = \pm 1$ is introduced for convenience. We assume, without loss of generality, $\sigma \ge 0$. Note that at $\sigma = 0$, \ie, $\phi = \phi_0$, one can choose either $\tau \ge 0$ or $\tau \le 0$. In what follows we assume $\xi\tau \geq 0$ at $\sigma = 0$. In these coordinates, the equation \threeRT\ now becomes
\eqn\SCC{-{1\over 2}{1\over (\sigma^2 + \tau^2)}\left(\partial^2_{\sigma} + \partial^2_{\tau}\right)\psi +{1\over 2}p(\tau^2 - \sigma^2)\psi - \gamma\psi = 0, \quad |\tau| \geq \chi\sigma, \quad \chi = 2 + \sqrt{3}.
}
The equation has decomposed into two parts. We note that
\eqn\SCCPP{-{1\over 2}\partial^2_{\sigma}\psi - {1\over 2}p\sigma^4\psi - \gamma\sigma^2 \psi = -\left(- {1\over 2}\partial^2_{\tau}\psi + {1\over 2}p\tau^4\psi - \gamma\tau^2 \psi \right), \quad |\tau| \geq \chi\sigma.
}
Therefore, this equation can naturally be solved using the separation of variables method. We write the wavefunction as a product of two functions as
\eqn\SEP{\psi(\tau, \sigma) = T(\tau)S(\sigma).
}
This ans\"{a}tz leads to the equation
\eqn\SEEQ{
 {S''\over S} + p\sigma^4 + 2\gamma \sigma^2 = -\left({T''\over T}  - p\tau^4 + 2\gamma\tau^2 \right), \quad  |\tau| \geq \chi \sigma.
}
Thus, for this equation to hold for all values of $\tau$ and $\sigma$ we need to demand
\eqn\POLZZSS{
\eqalign{
 -{d^2T\over d\tau^2} + (p \tau^4 - 2\gamma \tau^2 - l)T = 0,\cr
 -{d^2S\over d\sigma^2} - (p \sigma^4 + 2\gamma \sigma^2 - l)S = 0,
}
}
where $l$ is a constant and $|\tau| \geq \chi \sigma$. The separation constant $l$ is determined by imposing the appropriate boundary conditions on $T$ and $S$. Note that $S \equiv T(i\sigma)$. Thus, we only need to solve the first equation.

The conditions \bondWQR\ and \somDEF\ on the wavefunction are now given by
\eqn\CONDPS{ T(\pm\infty) = 0, \quad T(\tau) = -T(-\tau).
}

Consider the case $\tau > 0$. We write $T$ as
\eqn\SolvT{T(\tau) = e^{-a \tau^3 - b\tau}H(\tau), \quad a = {p^{1\over 2}\over 3} = {2^{3\over 4}\over 3}, \quad b = -{\gamma\over p^{1\over 2}} = -{\gamma\over 2^{3\over 4}}.
}
Plugging this into the equation for $T$ we get
\eqn\FORH{
{d^2H\over dz^2} - \left(3z^2 + \xi\right){dH\over dz} - \left(3z - \delta\right)H = 0,
}
where
\eqn\PARAH{z = \eta^{1\over 3}\tau,\quad \eta = 2a, \quad \xi = 2b/\eta^{1\over 3}, \quad \delta = \left(b^2 + l\right)/\eta^{2\over 3}.
}
The function $H(\delta, 0, \xi; z)$ is the triconfluent Heun function \Hortacsu. The triconfluent Heun function $H(\alpha, \beta, \nu; z)$ satisfies the equation 
\eqn\TriH{
{d^2H\over dz^2} - \left(3z^2 + \nu\right){dH\over dz} - \left((-\beta + 3) z - \alpha\right)H = 0.
}
Since \threeRT\ is the two dimensional generalization of \FouriRRJ, the triconfluent Heun function can be considered as the generalization of Airy function.

The solution for $T$ is then given by 
\eqn\SolT{
T(\tau) = \left\{\eqalign{
\exp\left[-\left({2^{3/ 4}\over 3}\tau^2 - 2^{-3/ 4}\gamma\right)\tau\right]H(\delta, 0, \xi; \eta^{1/ 3}\tau), & \quad \tau > 0, \cr
-\exp\left[\left({2^{3/ 4}\over 3}\tau^2 - 2^{-3/ 4}\gamma\right)\tau\right]H(\delta, 0, \xi; -\eta^{1/ 3}\tau), & \quad \tau < 0.
}\right.
}

We need to impose a boundary condition at $\phi = \phi_0$, \ie, at $\sigma = 0$. In general, $S(\sigma)$ has a definite parity. This corresponds to the following two possible boundary conditions. One boundary condition is
\eqn\BCsL{\left.S'(\sigma)\right|_{\sigma = 0} = 0.
}
The other boundary condition is
\eqn\BCsLL{\left.S(\sigma)\right|_{\sigma = 0} = 0.
}
However, we note that $\phi = \phi_0$ is the fixed point of the $\Z_2$ reflection symmetry mentioned above. Thus, at $\sigma = 0$, $S(\sigma)$ must vanish. That is, $S(\sigma)$ is an odd function. Therefore, $S(\sigma) = cT(i\sigma)$ for some constant $c$.

The wavefunction should be also continuous at $\phi - \phi_0 = -\pi/6$ and $\phi - \phi'_0 = \pi/6$ where $\phi'_0 = \phi_0 - \pi/3$. We next show that indeed it is continuous. Let $\psi^{\phi_0}_{\xi}(\tau, \sigma)$ denotes the wavefunction in the sector $\phi_0$ with $\xi = +1$ or $\xi = -1$. We thus have from \bondD\ that
\eqn\MATCGW{\psi^{\phi'_0}_{\xi'}(\tau, \sigma) = -\psi^{\phi_0}_{\xi}(\tau, \sigma).
}
We also have from \SEP\ and \SolT\ with $\xi = + 1$ that 
\eqn\WAVEFF{\psi^{\phi_0}_{+1}(\tau, \sigma) = \left\{\eqalign{
T(\tau)S(\sigma), & \quad \tau > \sigma > 0, \ \ie,\ 0  < \phi - \phi_0 <  \pi/6, \cr
-T(-\tau)S(\sigma), & \quad -\tau > \sigma > 0, \ \ie,\ -\pi/6 < \phi - \phi_0 < 0,
}\right.
}
where $T(0) = 0, \ T(\infty) = 0$ and $S(\sigma) = cT(i\sigma)$. Therefore, we observe that with $\xi' = -\xi$ the wavefunction is continuous.
\eqn\WAVEFFPP{\psi^{\phi'_0}_{-1}(\tau, \sigma) = \left\{\eqalign{
-T(-\tau)S(\sigma), & \quad -\tau > \sigma > 0, \ \ie,\ 0  < \phi - \phi'_0 <  \pi/6, \cr
T(\tau)S(\sigma), & \quad \tau > \sigma > 0, \ \ie,\ -\pi/6 < \phi - \phi'_0 < 0.
}\right.
}
This also implies the wavefunction $\psi\left(r, \phi \right)$ is even under the symmetry $\phi \to -\phi$. Therefore, the first derivative of $\psi$ w.r.t $\phi$ at $\phi = 0$ must vanish. That is,
\eqn\CONDVV{\left.\left(\tau\partial_{\sigma} - \sigma\partial_{\tau}\right)\psi_{+1}^{\phi_0}(\tau,\sigma)\right|_{\tau = \chi\sigma} = 0.
}
This further constrains the wavefunction. 

Comments:

 \item{(1)}The constraint \CONDVV\ ensures that the wavefunction matches smoothly across the boundaries of the different sectors. It should be viewed as a constraint on $l$. This will become evident as we go along. It is trivially satisfied at $\sigma = 0, \ \tau = 0$.
 \item{(2)}For small $\sigma$ and large $\tau$, \ie, near $\phi = \phi_0$ and far away from the origin $r \gg 1$, we have $\phi - \phi_0 = {\sigma/\tau}$ and $\tau^2/2 = r$. Interestingly, in this limit, the equation \POLZZSS\ reduces to
 \eqn\REDD{\left(-{1\over 2}{d^2\over dr^2} - {l\over 4r} + pr - \gamma\right)R(r) = 0,\quad S(\sigma) = c\cdot \sigma,
 } 
where $R(r) = r^{1/4}T(r)$ and $c$ is some constant. Note the appearance of the Cornell potential \refs{\Gottfried,\ \Eichten}. This is reasonable since it is known to describe heavy quarks. In fact, the first equation in \POLZZSS\ in general describes, interestingly, a particle in two dimensions in a Cornell potential. It reduces to
\eqn\ONcDDTT{\left({1\over 2}p_r^2 + {p^2_\phi\over 2r^2} - {l\over 4r} + pr - \gamma\right)\chi(r,\phi) = 0,
 } 
where $p_r$ and $p_\phi$ are the radial and angular momenta \threeev, and the $\phi$ dependence of $\chi(r, \phi)$ is $\exp(ik\phi)$; see appendix B for the details. The equation for $R$ reduces in the case $l = 0$ or strict large $r$ limit to Airy equation. The wavefunction is $\psi_0 \approx (\phi - \phi_0)\cdot r^{1/4}\cdot {\rm Ai}$. Thus, the spectrum is given by \statE\ and \FouriRRTHHY\ with $p = 2\sqrt{2}$
\eqn\SPECTHR{\bar\gamma^{(j,l)}_3 = \left(2\cdot 27\cdot \pi^2\right)^{1\over 3}\gamma = 3\left[3\pi^2\left(j + {1\over 2}\right) + \cdots\right]^{2\over 3} + \cdots.
}
where $j$ is large integer and $\cdots$ denotes corrections that involve $j$ and $l$. The factor 3 is due to the three flux tubes. In general we expect, at leading order, a factor of $n$. $n$ is the number of flux tubes or adjoint quarks. This is the case since the quarks are on a closed loop and each quark is connected to two flux tubes.

The spectrum $\gamma$ in general is determined by the boundary conditions at $\tau = 0$ and $\tau = \infty$, and the smoothness condition \CONDVV. We stress that the ans\"{a}tz \SEEQ\ and thus the conjecture holds provided we find non--trivial and real values for the spectrum $\gamma$ that are consistent with the boundary conditions and smoothness of the wavefunction \CONDVV. We hope to provide a detailed analysis of the spectrum $\gamma$ and related quantities in a future paper. We also hope to study the $n \geq 4$ cases in a separate paper in the future. 

The same analysis can be also done for the case where the fermions have different masses. We hope to study this and extend the discussion to higher orders in the coupling in the future.

In appendix A we provided representative plots of periodic motions in the associated classical system. 

\bigskip\bigskip

\noindent{\bf Acknowledgements:} I thank D. Kutasov for collaboration on the initial stages. I thank Igor Klebanov for useful comments on the draft and email correspondence.  I also thank Nava Gaddam for reading the draft and useful suggestions. This work is supported by the Department of Atomic Energy under project no. RTI4001.

\appendix{A}{3--parton Classical Dynamics}

The associated classical system to the $n$--body quantum system \FouriRRM\ is described by the Hamiltonian\foot{At the quantum level, $H\phi^{(m)}_n = E^{(m)}_n\phi^{(m)}_n,\ E^{(m)}_n = \bar\gamma^{(m)}_n/(2\pi^2)^{1/3}n$.}
\eqn\CLASSi{H = {1\over 2}\sum_{i = 1}^np_i^2 + V(q_1, \cdots, q_n), 
}
where the potential $V$ is given by
\eqn\CLASSicP{V(q_1, \cdots, q_n) = \sum_{i = 1}^n |q_i - q_{i + 1}|, \quad q_{n + 1} := q_1.
}

In this appendix we provide representative plots of classical periodic motions that possibly correspond bound states in the 3--parton quantum system. We will choose the center of mass position to be zero, thus $z_1(t) = 0$. The classical Hamiltonian in this case is given by
\eqn\ham{H = {1\over 2}p_2^2 +{1\over 2} p_3^2 + V( z_2, z_3), 
}
where the potential $V(z_2, z_3)$ is given by \VVTH.
The equations of motion are given by Hamilton's equations
\eqn\mot{{\dot z_2} = p_2, \quad {\dot z_3} = p_3, \quad -{\dot p_2} = {\partial H\over \partial z_2}, \quad -{\dot p_3} = {\partial H\over \partial z_3}.
}

In this classical system, there are two classes of closed periodic orbits, depending on initial conditions. We hope to discuss their semiclassical quantization in relation to the spectrum of the quantum system in a future work.

In the first class, the trajectories of the three quarks, \ie,\ $x_1(t),\ x_2(t)$ and $x_3(t)$, meet together only at zero position. In terms of $z_2(t)$ and $z_3(t)$, this implies, $z_2(t)$ and $z_3(t)$ meet or cross each other only at the origin. Thus, there is no exchange of momentum. The energies along the trajectories $z_2(t)$ and $z_3(t)$ are conserved independently.  A typical plot is given in Fig. 2.

\bigskip

\ifig\loc{All the masses are taken to be one in mass unit. On the left side we have the trajectories $x_1(t) {(\rm blue}),\ x_2(t) ({\rm orange})$ and $x_3(t) (\rm green)$. On the right side we have $z_2(t) = (x_2(t) - x_1(t))/\sqrt{2}$(blue) and $z_3(t) = \sqrt{3/2}x_3(t)$(orange). At $t = 0$, ${\dot x}_1 = (5/3){\dot x}_2, \ {\dot x}_2 = 1,\ {\dot x}_3 = -{\dot x}_1 - {\dot x}_2$. Here the plot is for a half period.}
\centerline{\epsfxsize5.4in\epsffile{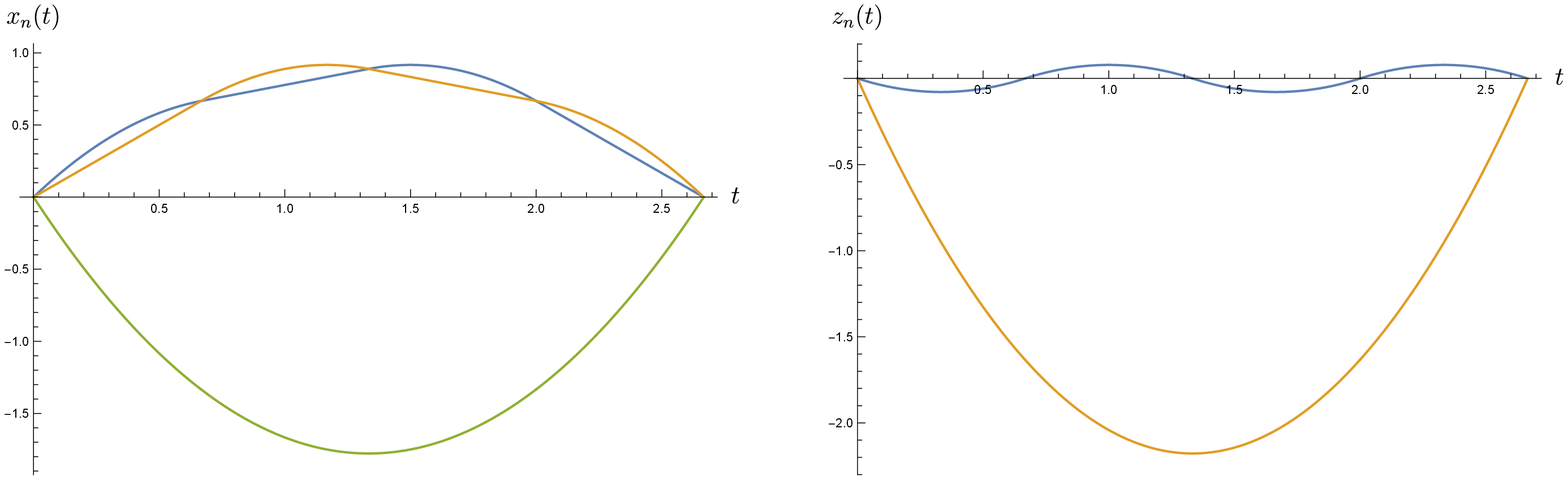}}

\bigskip

In the second class of closed periodic orbits, the trajectories $z_2(t)$ and $z_3(t)$ meet at least once away from zero position before they both meet again at the origin for the first non--zero time. In this case there is an exchange of momentum between $z_2$ and $z_3$. However, the total energy is conserved. A typical plot is given in Fig. 3.

\bigskip

\ifig\loc{All the masses are taken to be one in mass unit. On the left side we have the trajectories $x_1(t) ({\rm blue}),\ x_2(t) ({\rm orange})$ and $x_3(t) ({\rm green})$. On the right side we have $z_2(t) = (x_2(t) - x_1(t))/\sqrt{2}$(blue) and $z_3(t) = \sqrt{3/2}x_3(t)$(orange). At $t = 0$, ${\dot x}_1 = \alpha {\dot x}_2, \ {\dot x}_2 = 1,\ {\dot x}_3 = -{\dot x}_1 - {\dot x}_2, \ \alpha \approx 1.880810$. $\alpha$ is given by the real solution of a polynomial of degree 8. Here the plot is for a half period.}
\centerline{\epsfxsize5.4in\epsffile{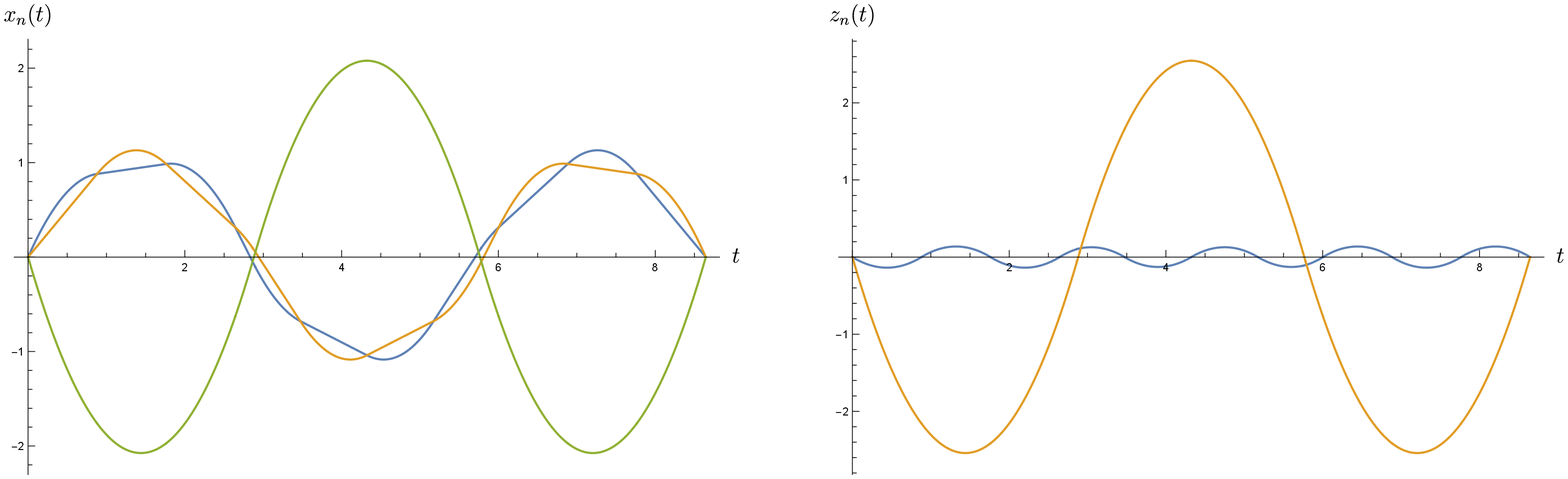}}

\bigskip

\appendix{B}{The Quartic Anharmonic Oscillator}

The quartic anharmonic oscillator is described by the equation \POLZZSS
\eqn\QAOQQ{
 -{d^2T\over d\tau^2} + (p \tau^4 - 2\gamma \tau^2 - q)T(\tau) = 0,  \quad p > 0,
}
here $q$ is the energy of the oscillator, $\gamma$ is the quadratic coupling and $p$ is the quartic coupling.

We first note that the equation 
\eqn\DUALEQQ{
 \left({1\over 2}p_r^2 + {p^2_\phi\over 2r} + pr - \gamma\right)\psi(r, \phi) = 0,
 }
 where
 \eqn\MOMDTH{ p^2_r = -{1\over r^{2l}}{\partial\over\partial r}r^{2l}{\partial\over\partial r} = -{1\over r^{l}}{\partial^2\over \partial r^2}r^{l} + l\left(l - 1\right){1\over r^2}, \quad p_\phi^2 = -{\partial^2\over \partial \phi^2},
}
is equivalent to \QAOQQ\ with $l = 1/ 4$. To see this we write
\eqn\EQUIV{
r = {1\over 2}\tau^2, \quad \psi(r, \phi) = T(r)e^{\sqrt{q\over 2}\phi}.
} 
This gives precisely \QAOQQ. 

We write \DUALEQQ\  as
 \eqn\ONcDD{\left(-{1\over 2}{d^2\over dr^2} + {l(l - 1)\over 2r^2} - {q\over 4r} + pr - \gamma\right)R(r) = 0,
 } 
where $l = 1/4$ and $R(r) = r^{l}T(r) = r^{1/4}T(r)$. As it is clear from \MOMDTH, the $1/r^2$ term comes from the radial momentum operator. We note that \ONcDD\ can be put in the form
\eqn\ONcDDTT{\left({1\over 2}p_r^2 + {l^2\over 2r^2} - {q\over 4r} + pr - \gamma\right)\chi(r) = 0,
 } 
where $\chi(r) = r^{-1/2}R(r)$, $p_r$ is the two dimensional radial momentum operator \threeev\ and $l $ can be thought of as the orbital angular momentum. Therefore, the quartic oscillator \POLZZSS\ equivalently describes a particle in two dimensions in a Cornell potential,
\eqn\CORRR{V(r) = - {q\over 4r} + pr.
}

In the case $q \neq 0$, the large $r$ limit of \ONcDD\ gives
\eqn\ONcDDHYME{\left(-{1\over 2}{d^2\over dr^2} - {q\over 4r} + pr  - \gamma\right)R(r) = 0.
} 
Note we kept the $1/r$ term to account for the $q$ dependence.

We also note that for small $r$ \ONcDD\ is the (radial part of the) hydrogen problem. In this limit the equation reduces to 
\eqn\ONcDDHY{\left(-{1\over 2}{d^2\over dr^2} - {q\over 4r} + {l(l - 1)\over 2r^2}  - \gamma\right)R(r) = 0.
} 
Here the coupling $\gamma$ is the energy of the particle and the energy $q$ measures its charge. The solutions are given by Whittaker functions.

Note also \ONcDDHY\ is equivalent to setting $p = 0$ in \QAOQQ. Thus, it also describes the harmonic oscillator. 

We note that in the case in which $q = 0$ the equation \ONcDD\ reduces to 
\eqn\ONcDDHY{\left(-{1\over 2}{d^2\over dr^2} + {l(l - 1)\over 2r^2} + pr  - \gamma\right)R(r) = 0.
} 
The solutions give a relation between the couplings $p$ and $\gamma$ which corresponds to a state in \QAOQQ\ with zero energy. A semiclassical calculation \Cornwall\ gives,
\eqn\GAGAGA{\gamma_{(n, l)} = \left[{3\over 4}\pi\cdot p\left(n + {l\over 2}\right)\right]^{2\over 3}. 
}

In the large $r$ limit the equation \ONcDDHY\ further reduces to
\eqn\ONcDDAIR{\left(-{1\over 2}{d^2\over dr^2} +pr  - \gamma\right)R(r) = 0.
} 
In this limit the solutions are given by Airy functions. $\gamma$ is given by \GAGAGA\ with $l = 0$ and large $n$.

\bigskip

\listrefs

\bye